\documentclass[twocolumn]{aa} 

\usepackage{graphicx}
\usepackage{gensymb}
\usepackage{txfonts}
\usepackage{natbib}
\usepackage{float}
\usepackage{mathtools}
\usepackage[english]{babel}
\bibpunct{(}{)}{;}{a}{}{,}    
\usepackage{cancel}
\usepackage{xcolor}
\usepackage[colorlinks=true,citecolor=blue,linkcolor=blue]{hyperref}

\newcommand{\rotrate}{$\Omega_{\odot}$}

\newcommand{\rvar}{$R_{var}$}
\newcommand{\K}{\textit{Kepler}}

\defcitealias{Nina1}{N20b}
\defcitealias{McQuillan2014}{McQ14}
\defcitealias{Santos2021}{S21}
\defcitealias{Isik2018}{Paper I}

\begin{document}

   \title{Forward modelling of brightness variations in Sun-like stars}

   \subtitle{II. Light curves and variability}
\titlerunning{Forward modelling of brightness variations in Sun-like stars II}
\authorrunning{N.-E. Nèmec et al.}
\author{N.-E. Nèmec,
          \inst{1,2}
          A. I. Shapiro\inst{2},
          E. I\c{s}{\i}k\inst{3},
          S. K. Solanki\inst{2,4}, and
          T. Reinhold\inst{2}}

   \institute{Institut für Astrophysik und Geophysik,Georg-August-Universität Göttingen, Friedrich-Hund-Platz 1, D-37077 Göttingen, Germany\\
    \email{nina-elisabeth.nemec@uni-goettingen.de}
    \and
   Max-Planck-Institut für Sonnensystemforschung, Justus-von-Liebig-Weg 3, D-37077 Göttingen, Germany
             \and 
             Department of Computer Science, Turkish-German University, \c{S}ahinkaya Cd. 94, 34820 Beykoz, Istanbul, Turkey
            \and 
            School of Space Research, Kyung Hee University, Yongin, Gyeonggi, 446-701, Korea
             }

   \date{}

 
  \abstract
   {
   The amplitude and morphology of light curves of solar-like stars change substantially with increasing rotation rate: 
   brightness variations get amplified and become more regular, 
  which has so far not been explained. 
   }
  {We develop a modelling approach for calculating brightness variations of stars with various rotation rates and use it to explain observed trends in stellar photometric variability.
  }
   {We combine numerical simulations of magnetic Flux Emergence And Transport (FEAT) with a model for stellar brightness variability to calculate synthetic light curves of stars as observed by the \K{} telescope. We compute the distribution of magnetic flux on the stellar surface for various rotation rates and degrees of active-region nesting (i.e., the tendency of active regions to emerge in the vicinity of recently emerged ones). Using the resulting maps of the magnetic flux, we compute the rotational variability of our simulated stellar light curves as a function of rotation rate and nesting of magnetic features and compare our calculations to \K{} observations.}
   {We show that both rotation rate and degree of nesting have a strong impact on the amplitude and morphology of stellar light curves. In order to explain the variability of the bulk of \K{} targets with known rotation rates, we need to increase the degree of nesting to values much larger than on the Sun.}
   {The suggested increase of nesting with the rotation rate can provide clues to the flux emergence process for high levels of stellar activity. }
   



   \maketitle
%
\section{Introduction}

 Planet hunting missions such as the Convection, Rotation and planetary Transits \citep[CoRoT][]{COROT,COROT2}, \textit{Kepler} \citep{KEPLER} and the Transiting Exoplanet Survey Satellite \citep[TESS,][]{TESS} allow studying stellar brightness variations caused by transits of magnetic features as stars rotate. Such brightness variations were discovered for the Sun almost half a century ago \citep{Willsonetal1981, WillsonandHudson1981}. Since then the models of solar brightness variations have matured and are now not only capable of accurately reproducing most of the available measurements \citep[see][for reviews]{Solanki2013, Ermolli2013} but they also provide  a starting point for explaining a plethora of stellar photometric data \citep[see, e.g.][]{Lagrange2010, Meunier2013, Meunier2015, Borgniet2015, Nina1}.
 
Recently \cite{Nina1} (hereafter \citetalias{Nina1}) have combined the Spectral And Total Irradiance Reconstruction model \citep[SATIRE,][]{Fligge2000,Krivova2003} together with a surface flux transport model \citep[SFTM,][]{Cameron2010} to compute the power spectra of solar brightness variations  as they would be measured at different inclinations, i.e. the angle between solar rotation axis and direction to the observer. These calculations helped to remove a number of important observational biases when comparing solar variability to that of other stars \citep{Nina2,Timo2020}. Notably, by employing the \citetalias{Nina1} model and using the approach developed by \cite{Witzke2018, Witzke2020} to extend it to stars with non-solar metallicities, \cite{Timo2020_2} have found that rotation periods of a majority of the G-dwarfs with near-solar age remain undetected. These results provided an explanation for the discrepancy between the predictions of the number of Sun-like rotators in the \textit{Kepler} field and the actual number of detected ones \citep[see][]{VanSaders2019} .
 
In this work, we make an additional important extension of the solar paradigm for modelling variability of stars rotating faster than the Sun. Namely, we combine the solar variability model of \citetalias{Nina1}, which was extensively tested against solar irradiance measurements, with the modelling framework for computing the surface distribution of magnetic flux on stars with solar fundamental parameters but various rotation rates developed by  \cite{Isik2018} (hereafter \citetalias{Isik2018}). The Flux Emergence And Transport (FEAT) model presented in
 \citetalias{Isik2018} involves physics-based calculations of the emergence latitudes and tilt angles of bipolar magnetic regions (BMRs) for given stellar rotation rates, and the subsequent modelling of the evolution of the radial magnetic flux at the photosphere via an SFTM. The FEAT model is self-consistently able to reproduce the observations of polar spots that appear on stars with rotation periods below about 3 days \citep[see, e.g.,][]{Jeffers2002,Marsden2004,Jarvinen2006,Waite2015}. Recently, the FEAT model was successfully applied to the young solar analogue EK Dra \citep{Senavci21}, to explain the Doppler images that indicated 
near-polar spots and extended spot patterns towards low latitudes. In the present work, we extend the model of \citetalias{Isik2018} to calculate brightness variations of stars with a variety of rotation rates observed at various inclination. We also allow for different degrees and modes of nesting of magnetic features on their surfaces (i.e. the tendency of active regions to emerge in the vicinity of recently emerged regions). We compare our results to the observational trends found by \cite{McQuillan2014}  and \cite{Santos2021} from {\it Kepler} data and propose a possible explanation of these trends.

In Sec.~\ref{model} we briefly describe the FEAT model  \citep[see][for a more detailed description]{Isik2018} and explain how this model is extended for  calculating stellar brightness variability. In Sec.~\ref{lightcurves} we discuss the resulting light curves (LCs). These LCs are then used to calculate the amplitude of the variability, which we compare to observations in Sec.~\ref{obs}. We discuss our findings within the frameworks of various other recent studies in Sec.~\ref{discussion}, before we present our conclusions in Sec.~\ref{conclusions}.

\section{Model}\label{model}

Our model consists of two building blocks: calculations of the surface distribution of magnetic features and the subsequent calculations of their effect on stellar brightness. 

\subsection{Flux emergence and transport}\label{FEAT}

To simulate the magnetic flux emergence on other stars, we utilise the FEAT model \citepalias[][]{Isik2018}. 

In essence, the FEAT model extends the pattern of emergence and evolution of the magnetic fields observed on the surface of the Sun to stars rotating faster than the Sun (and, thus, more active than the Sun). This is done in five steps \citepalias[see][Fig.~B1]{Isik2018}:
\begin{description}
\item[Step 1.] We adopt the synthetic record of solar active-region emergences during cycle 22 from \cite{Jiang2011_1}. While this approach cannot be used to reconstruct the solar irradiance on a specific day,  \citetalias{Nina1} have shown that \cite{Jiang2011_1}  records allow reproducing the overall pattern of solar variability. Cycle 22 was chosen, as it represents a cycle of moderate to strong activity level, making it suitable for modelling the most active stars.
\item[Step 2.] We define the time-dependent emergence rate of BMRs on a star as $S_{\star}(t)= S_{\odot}(t) \cdot \tilde{s}  $, where $S_{\star}$ and $S_{\odot}$ are stellar and solar emergence rates, respectively, and $\tilde{s}$ is a scaling factor. To reflect the observed rotation--activity relation, we followed \citetalias{Isik2018} and took $\tilde{s}= \tilde{\omega} \equiv \Omega_{\star}/\Omega_{\odot}$, where $\Omega_{\star}$ is the rotation rate of a star and $\Omega_{\odot}$ is the solar rotation rate. 
\item[Step 3.] The resulting input record of emergences  is mapped down to the base of the convection zone using thin flux tube simulations \citep{Schuessler1996,Isik2018}  for the solar rotation rate \rotrate{}. 
\item[Step 4.] The record of emergences at the base of the convective zone is mapped back to the surface, but in contrast to Step 3 using thin flux tube simulations for a star  
with a given rotation rate $\Omega_\star$. These simulations follow the rise of the flux tubes throughout the convection zone up to the surface of the star, where they emerge in the form of a loop with two footpoints of opposite polarity (i.e. BMRs). An important feature of these  simulations is that they account for the Coriolis effect that pushes rising flux tubes towards higher latitudes for higher rotation rates \cite{Schuessler1992}.
\item[Step 5.] Assuming that the size distribution of BMRs does not change with the rotation rate, we modify the emergence locations of flux loops, to simulate different degrees and modes of active-region nesting \citep[see a similar approach by][]{Isik2020}, motivated by the observed activity complexes on the Sun. The details of this step will be described later in this Section.
\item[Step 6.] In this last step, the calculated locations of emergence and the tilt angles of BMRs are fed into the surface flux transport model (SFTM).  The SFTM describes the passive transport of magnetic fields of the BMRs on the surface of stars, by taking into account both large-scale flows (meridional flow and differential rotation) and the diffusion of the fields \citep{Cameron2010, Isik2018}.
\end{description}

\begin{figure}[!h]
\centering
\includegraphics[width=0.45\columnwidth]{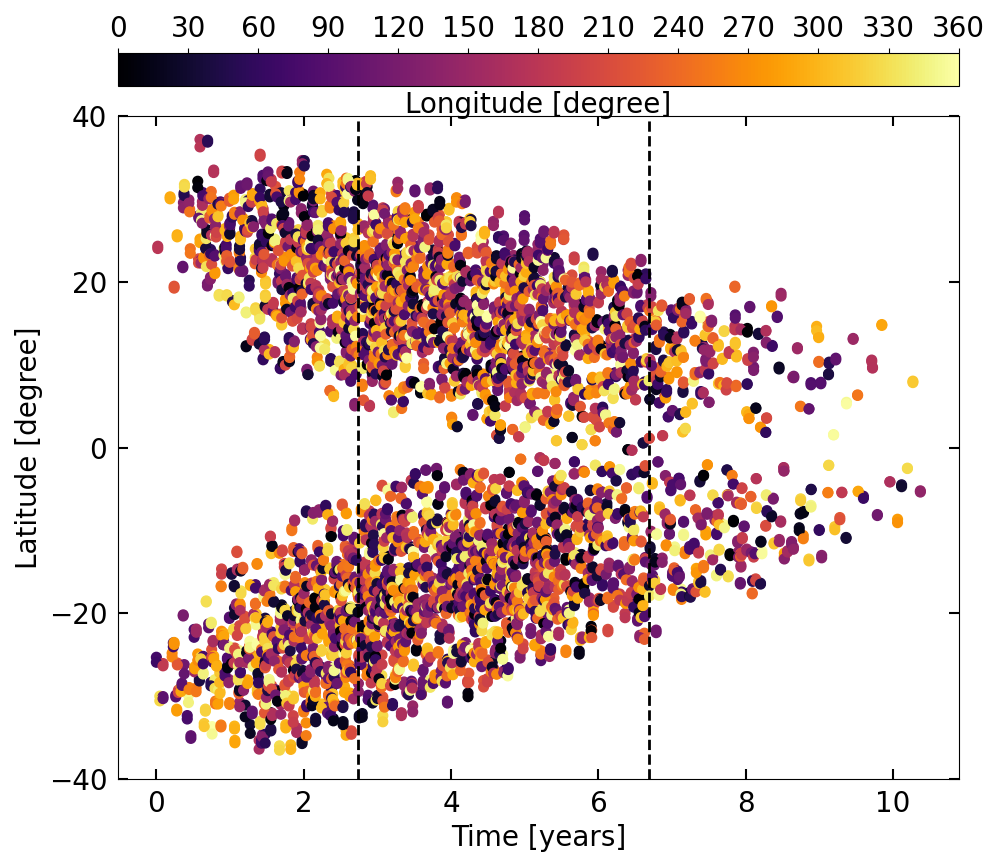}\quad
\includegraphics[width=0.45\columnwidth]{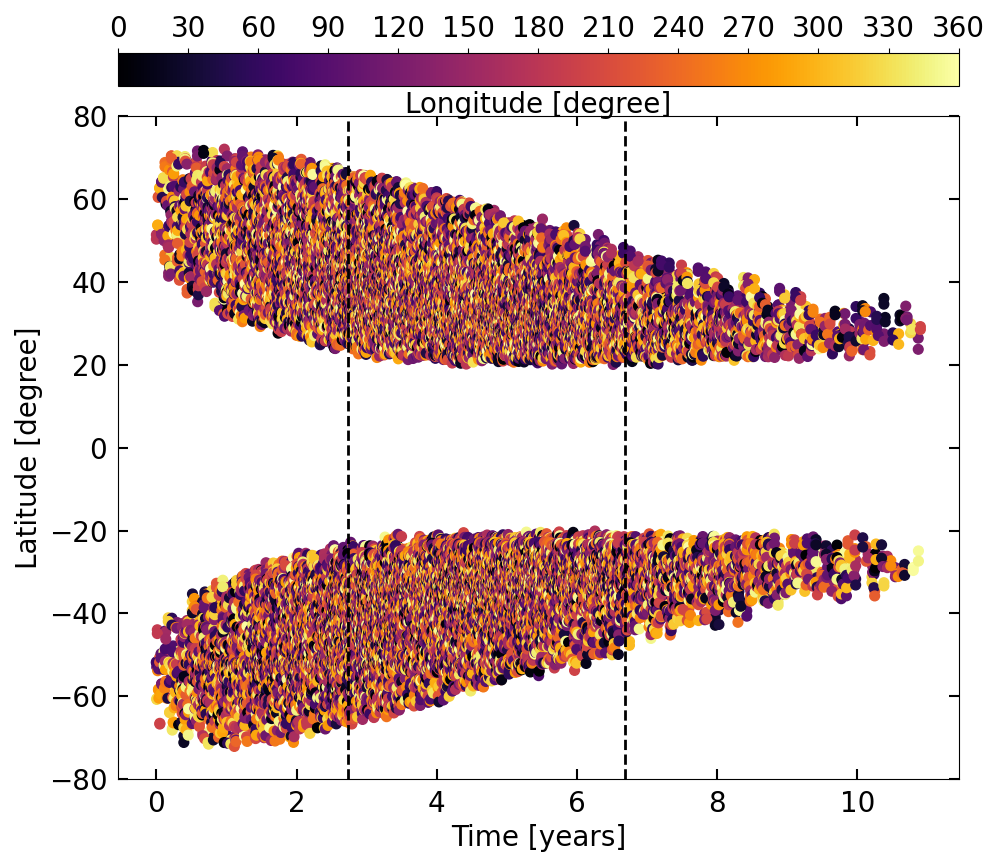}\quad
\caption{Butterfly diagrams of BMR emergence for 1\rotrate{} left panel and 8\rotrate{} right panel. The colour-bar gives the longitudes of emergence and the vertical dashed lines indicate the 4 years of the 11 year cycle considered in our brightness variation calculations. No amount or nesting is included here.}
\label{fig:butterfly_rots}
\end{figure}

In the present work, we employ the steps outlined to obtain the surface distribution for stars with 1, 2, 4, and 8\rotrate{}. The emergence patterns of the BMRs (e.g. the input record) for 1\rotrate{}  and 8\rotrate{} are shown in Fig.~\ref{fig:butterfly_rots}. 
The main purpose of this study is to model stellar variability on the rotational timescale. We follow the approach of \citetalias{Isik2018} and take the solar temporal profile of activity (even though it is unlikely to be observed on the faster rotating and more active stars, we discuss this choice later in this paper), but we model the light curves only over the four-year window centred at the activity maximum (marked in Fig.~\ref{fig:butterfly_rots}). Our light curves (and the resulting variability amplitudes) correspond to a representative stellar activity level scaled from the solar maximum. Hence they are largely unaffected by the temporal profile of the underlying activity cycle.

\begin{figure}
    \centering
    \includegraphics[width=\columnwidth]{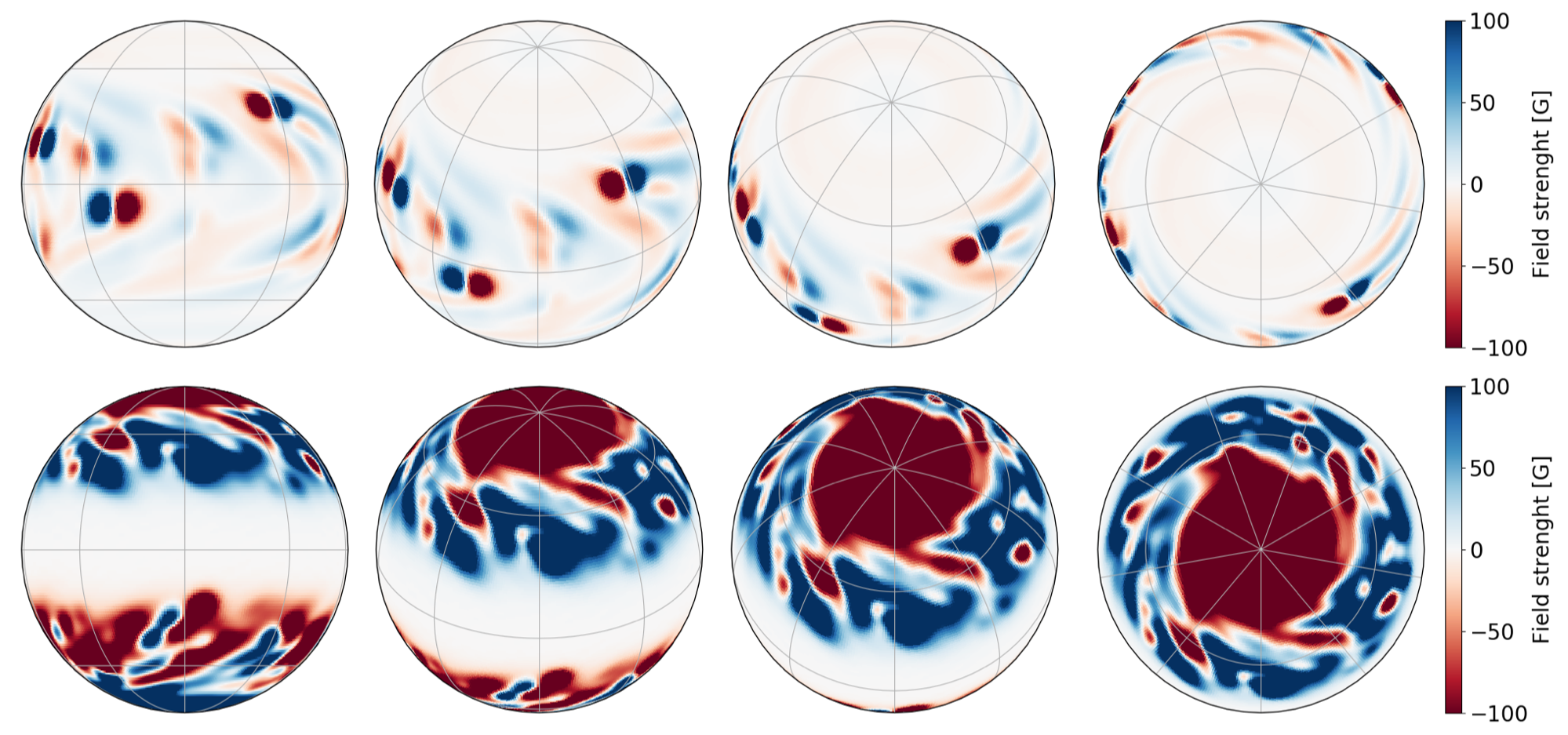}
    \caption{{\bf Snapshot of} magnetic field distribution for 1\rotrate{} (top row) and 8\rotrate{} (bottom row) as returned by the FEAT model at various inclination angles around maximum of the activity cycle. First column shows $i=90$\degree, second column $i=57$\degree,  third column $i=30$\degree, and forth column $i=0$\degree.}
\label{fig:fields}
\end{figure}

In Fig.~\ref{fig:fields} we present snapshots of the magnetic field distribution on the surface of stars with 1\rotrate{} (top row) and 8\rotrate{} (bottom row) for various inclinations. The figure nicely demonstrates the difference in the emergence frequency, as the 8\rotrate{} star is covered by more BMRs. Also the difference in the latitudinal distribution is remarkable. This is a result of Step 4 of the FEAT approach, which in turn leads to the formation of strong polar fields for the fast rotator depicted here.
We also show snapshots of the magnetic field distribution of the surface of stars with \rotrate{} and 8\rotrate{} for various inclinations in Fig.~\ref{fig:fields}. 

Additionally, the emergence pattern of BMRs is altered by introducing active-region nesting. 
This effect is observed on the Sun, albeit to a small degree \citep[see, e.g.,][who estimated that 40--50\% of solar active regions can be associated to nests]{Pojoga2002} and it was recently suggested that nesting can be substantially stronger on highly variable stars with near solar rotation rates \citep{Isik2020}.
Following \cite{Isik2020}, we introduce two modes of nesting (Step 5): the active-longitude (AL) and the free nesting (FN) modes. 
In both modes, we define a degree of nesting, $p$, which gives the probability of a given BMR to be part of a nest. In the AL mode, those nests are centered around two fixed longitudes with a separation of 180 degrees.
If the BMR is drawn to belong to a nest, then it is shifted to one of the two ALs (with equal probability). Its new longitude is defined by  a 1D normal distribution around the AL  (with the standard deviation $10^\circ$), whereas the latitude is kept unchanged. This ensures that the latitudes of emergence still follow the general trends of the solar butterfly diagram. In the FN mode, nesting occurs around central latitudes and longitudes, which are randomly picked from the unaltered (i.e. non-nested) emergence record. If the BMR is drawn to belong to a nest then its emergence location is moved to a random location drawn from a 2D normal distribution with the mean at the nest centre having standard deviations of $2^\circ$ in latitude and $3^\circ$ in longitude. For more details on the nesting definition and choice of the parameters we refer to \citealt{Isik2018} for the FN employed here and \citealt{Isik2020} for AL. 

\begin{figure}[!h]
\centering
\includegraphics[width=.48\columnwidth]{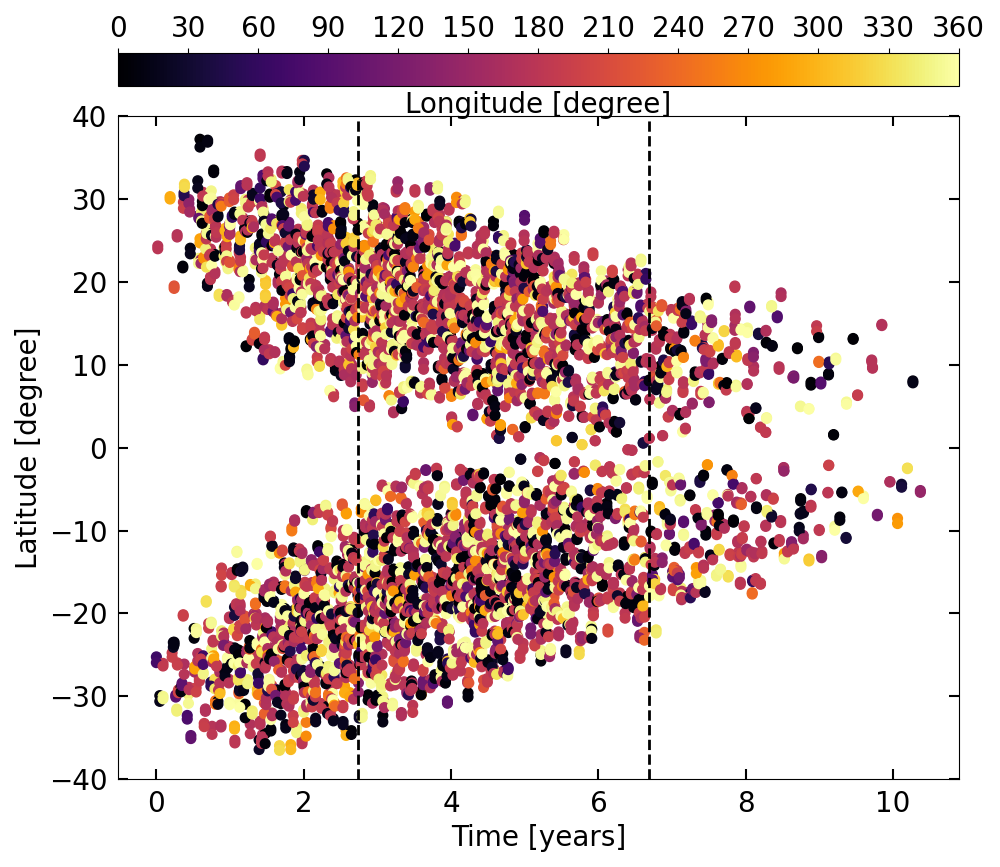}\quad
\includegraphics[width=.48\columnwidth]{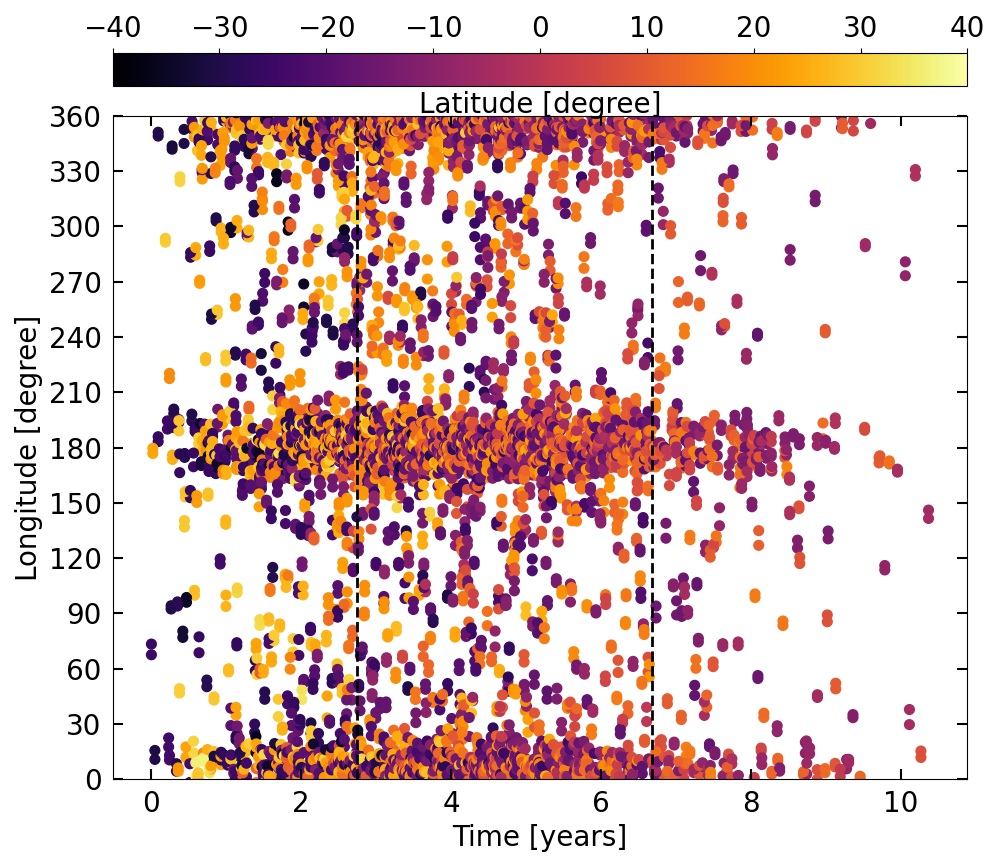}
\medskip
\includegraphics[width=.48\columnwidth]{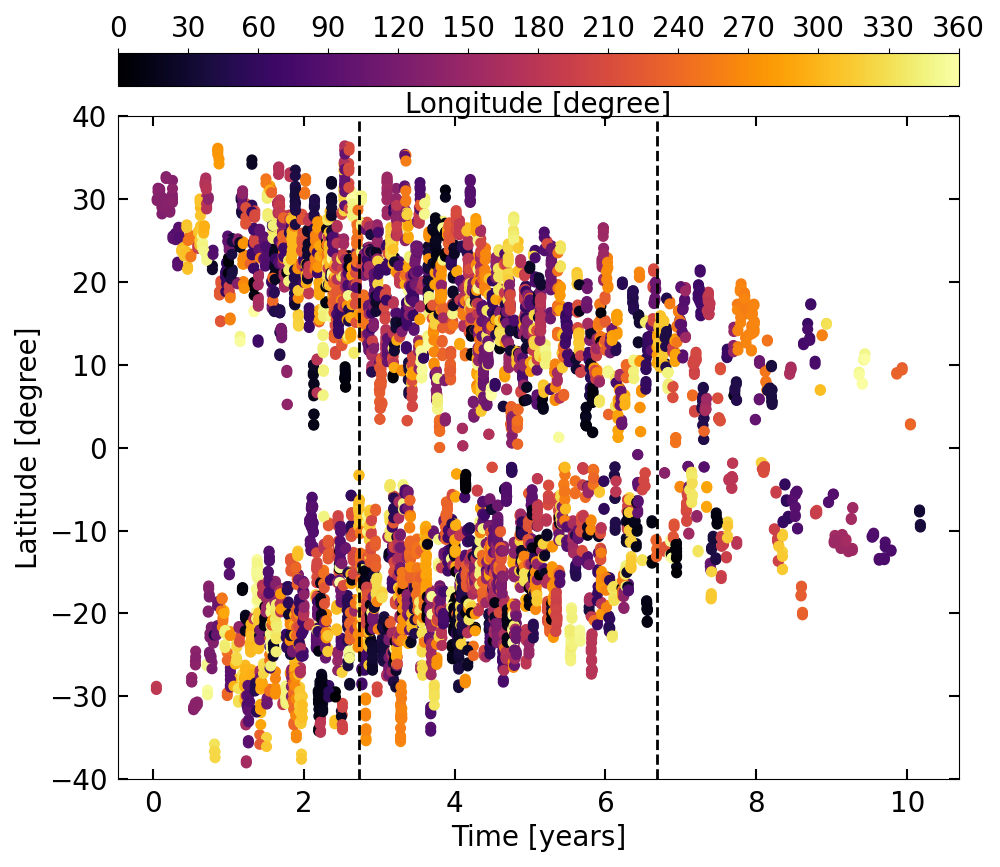}\quad
\includegraphics[width=.48\columnwidth]{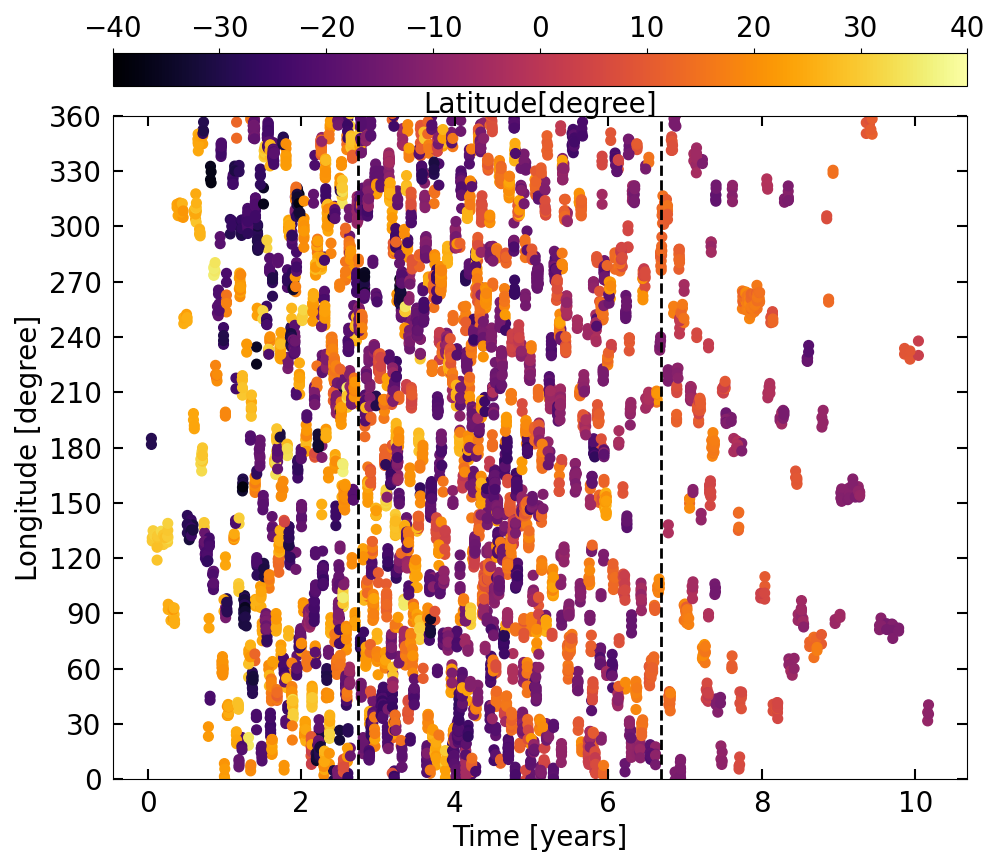}
\caption{Butterfly diagrams of BMR emergence for 1\rotrate{}, with $p=0.7$ in the different nesting modes considered in this work. Top row: AL nesting, bottom row: FN nesting. Left panels are time-latitude diagrams with the colourbar indicating the longitudes, right panels are time-longitude diagrams, with the colourbar indicating the latitude.}
\label{fig:butterfly_solrot}
\end{figure}

We show a comparison  between the two modes of nesting for {1\rotrate} in Fig.~\ref{fig:butterfly_solrot}. In order to distinguish more easily between the two modes, we show both time-latitude (left panels) and time-longitude diagrams, with the colourbar indicating the longitude and latitude each. The AL nesting is easier to identify in the time-longitude diagrams, whereas the FN nesting is easier to observe in a time-latitude graph.

\subsection{Defining the area coverages}\label{filling_factors}

The output of the FEAT model consists of full stellar surface maps  (360$\degree$ by 180$\degree$) of  magnetic fields, with a resolution of 1$\degree\times 1\degree$ per pixel. As our brightness calculations rely on the area coverages of spots and faculae, we have to first convert the magnetic field maps into the surface distributions of  spots and faculae. \citetalias{Nina1} did this by following the evolution of sunspots after their emergence to calculate the coverage of the solar disk by spots. While this approach proved itself to be accurate for calculating solar variability \citep[see also][]{Dasi2014} it did not account for spots which appeared due to the superposition of the magnetic flux from different active regions. In other words, the main assumption of this approach was that all spots on the stellar surface have emerged {\it as spots} within the corresponding active region. While this is a good assumption for the present Sun it prohibits the formation of polar spots, which are observed on young and rapidly rotating G stars \citep{Jeffers2002,Marsden2004,Jarvinen2008,Waite2015,Senavci21}. These polar spots most likely form via flux superposition. We therefore take a slightly different approach than in  \citetalias{Nina1}. We acknowledge that, apart from the intrinsic mechanism in the FEAT model it is theoretically also possible to form polar spots by overly increasing the rate of BMR emergence \citep{Schrijver2001}, or meridional flow rate \citep{Holzwarth2006}, but leaving the emergence latitudes solar-like. 

To calculate the  spot area coverage of a given pixel of the synthetic magnetogram returned by FEAT we define two thresholds: a lower cut-off, $B_{min}$, and an upper saturation level, $B_{max}$. The spot coverage of a given pixel is related to the field in the pixel following
\begin{equation}
    \alpha_{s}^{m,n}=
  \begin{cases}
    0 \quad \text{if} \quad |B_{mn}|< B_{min} \\
	 \frac{|B_{mn}|}{(B_{max}-B_{min})} -\frac{B_{min}}{(B_{max}-B_{min})} \quad \text{if} \quad B_{min} <=|B_{mn}| < B_{max} \\
	1 \quad \text{if}  \quad |B_{mn}| >= B_{max},
	\end{cases}
	\label{eq:filling_factors}
\end{equation} 
\noindent where $\alpha_{s}^{m,n}$ is the spot filling factor of a  pixel with coordinates m and n of the magnetogram, and $|B_{mn}|$ is the absolute value of the field in this pixel. 

In order to calculate the faculae area coverage in each pixel, we follow an  approach similar to \citetalias{Nina1} by setting a saturation threshold $B_{sat}$ \citep[see also papers describing SATIRE, e.g.,][]{Fligge2000,Krivova2003}. If we find that a pixel is covered already partially by spots, we disregard this pixel for the facular masking. If a given pixel is spot-free, we calculate the faculae area coverage  following:

\begin{equation}
    \alpha_{f}^{m,n}=
  \begin{cases}
  
  \frac{|B_{mn}|}{|B_{sat}|} \quad if \quad |B_{mn}| < B_{sat} \\
   1 \quad if \quad |B_{mn}| >= B_{sat}, 
	\end{cases}
	\label{eq:filling_factors_2}
\end{equation}

The values of the parameters are selected such that the current model returns  the same rotational variability (see Sect.~\ref{params}) for the solar case (i.e. $\Omega_{\star}=\Omega_{\odot}$) as the \citetalias{Nina1} model.

\subsection{Calculating the brightness variations}\label{bright_var}

Using Eq.~\ref{eq:filling_factors} and \ref{eq:filling_factors_2} we obtain maps of area coverages per pixel of the visible stellar disc. What part of the full surface (hence the disc) of a star is visible depends on the inclination between the stellar rotation axis and the line-of-sight to the observer. The spectral irradiance, $S(t,\lambda)$ (i.e., the spectral stellar flux, normalized to 1 AU), where $t$ is the time and $\lambda$ the wavelength, is then calculated  by summing up the intensities weighted by the corresponding  area coverages by magnetic features of a pixel as given by 

\begin{equation}
    S(t,\lambda) = S^{q}(\lambda) +\sum_{mn} \sum_{k}   \left (I_{mn}^{k}(\lambda)-I_{mn}^{q}(\lambda) \right ) \, \alpha_{mn}^{k}(t) \,
    \Delta\Omega_{mn}.
\end{equation}
\noindent 
The summation is done over the pixels of the maps of the visible 2D stellar disc and the $m$ and $n$ indices are the pixel coordinates (longitude and latitude, respectively), $\alpha_{mn}^{k}$ is the pixel ($m$,$n$) coverage by magnetic feature $k$ (in the present work faculae, umbra or penumbra), $\Delta \Omega_{mn}$ is the solid angle of the area on the stellar disc corresponding to one pixel, as seen from the distance of 1 AU. $I_{mn}^{k}$ is the intensity spectrum of magnetic feature $k$ observed at the location corresponding to pixel ($m$,$n$). 
We use the values computed by  \cite{Unruh1999}
with the radiative transfer code ATLAS9  \citep{Kurucz1992, Castelli1994}.  

S$^{q}$ is the quiet--star irradiance, defined as
\begin{equation}
   S^{q}(\lambda_w) = \sum_{mn} I_{mn}^{q}(\lambda_w)\Delta\Omega_{mn}.
\end{equation}
\noindent 
Note that the solid angles of the pixels, as well as the corresponding intensity values depend on the vantage point. Hence $S(t,\lambda)$  is sensitive to the stellar inclination. The calculations presented in this work are performed in the \textit{Kepler} passband, following 
\begin{equation}
LC(t) =  \int\limits_{\lambda_1}^{\lambda_2} R(\lambda) S(\lambda,t)  \frac{\lambda}{hc} \, d\lambda,
\end{equation}
\noindent where $\lambda_1$ and $\lambda_2$ are the blue and red threshold wavelengths of the filter passband, $R(\lambda)$ is the response function of the filter and $S(\lambda,t)$ is the spectral irradiance at a given wavelength and time $t$, $h$ is the Planck constant, and $c$ is the speed of light.

\subsection{Defining the parameters of the model}\label{params}


As mentioned previously, we fix the parameters of the model introduced in Sect.~\ref{filling_factors} to return the same level of rotational variability, represented via \rvar{}, of the solar case as \citetalias{Nina1}. \citetalias{Nina1} showed that their model is able to reproduce the observed  brightness variations of the Sun. 
The \citetalias{Nina1} model and the model we develop here are both based on the SFTM, with similar underlying statistical BMR emergence records. We therefore use the \citetalias{Nina1} model as our reference in the present work. For this,  we considered the four-year interval (indicated by vertical dashed black lines in Fig.~\ref{fig:butterfly_rots}) around the maximum of the synthetic cycle.


We then split the time series into 90-day segments, de-trended the new time series by their mean value, and  calculated the difference between the extrema in each of the segments using the approach outlined above and that of \citetalias{Nina1}. We note that  we directly consider the difference between the extrema instead of the differences between the 95th and 5th percentiles of sorted flux values, as is usually done in the literature with the more noisy \textit{Kepler} measurements  \citep[see, e.g.,][]{Basri2013}. 

\begin{figure}
\centering
\includegraphics[width=0.75\columnwidth]{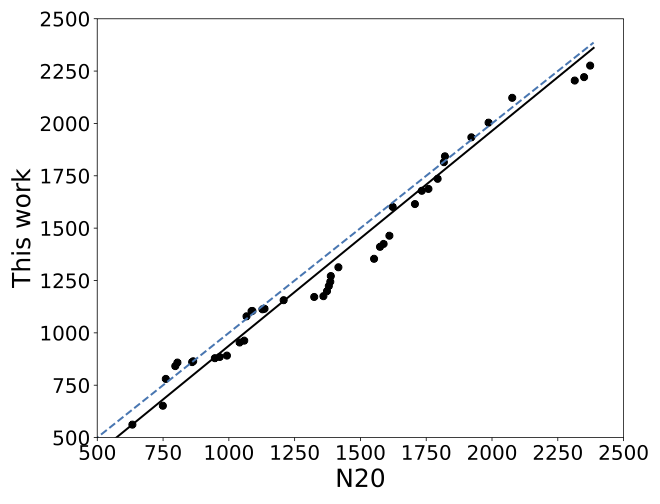}
\caption{Comparison of $R_{var}$ (in ppm) as returned by the \citetalias{Nina1} model and the present model. All values are given in ppm. Black solid line gives the linear regression, whereas the blue dashed line is the 1-to-1 correspondence between the two models}
\label{fig:params}
\end{figure}

We show a comparison between the \rvar{} values for the solar case as returned by \citetalias{Nina1} and the present model in Fig.~\ref{fig:params}. The best-values for the parameters $B_{min}$, $B_{max}$, and $B_{sat}$ were found to be 60, 700 and 250 G respectively and were chosen, as they resulted in a slope of the linear regression close to unity (1.029) and a high $r^2$ value (0.957). The mean rotational variability in the present model is comparable to that of the \citetalias{Nina1} model (1459.6 to 1457.71 ppm). However, we note that the threshold approach used in this model favours spots over faculae. While this is not necessarily accurate for the solar rotator in the present work, it leads to the formation of polar spots and spot domination of the stellar variability for the faster rotating stars. 

\begin{figure}
\centering
\includegraphics[width=\columnwidth]{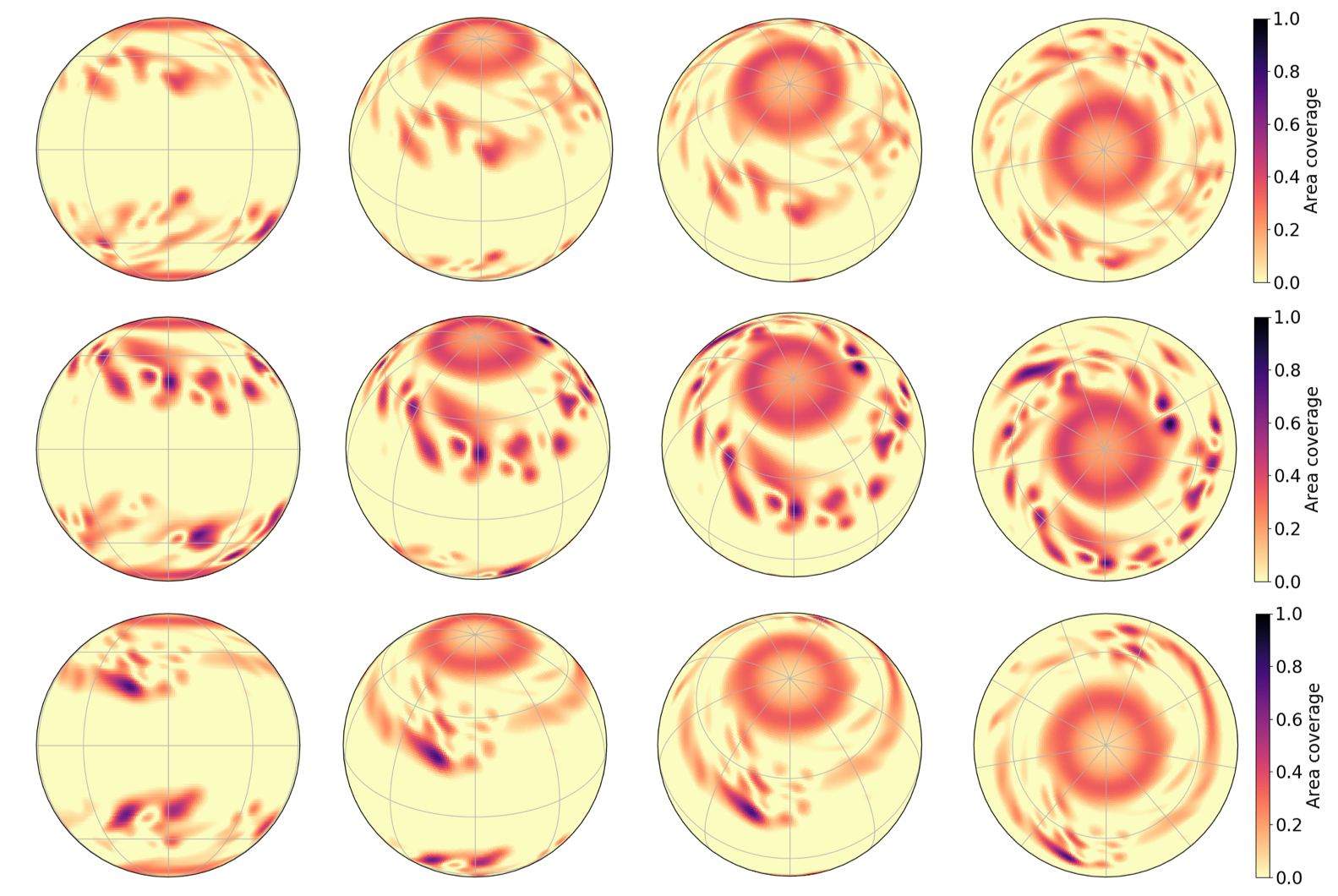}
\caption{Spot area coverage per pixel for different nesting realisations at different viewing angles for the 8\rotrate{} case. First column shows $i=90$\degree, second column $i=57$\degree,  third column $i=30$\degree, and forth column $i=0$\degree. Top row is the non-nested case, middle row includes $p=0.7$ in the free-nesting (FN) case, bottom row includes $p=0.7$ in the active-longitude (AL) case.}
\label{fig:map_spots}
\end{figure}

\begin{figure}
\centering
\includegraphics[width=\columnwidth]{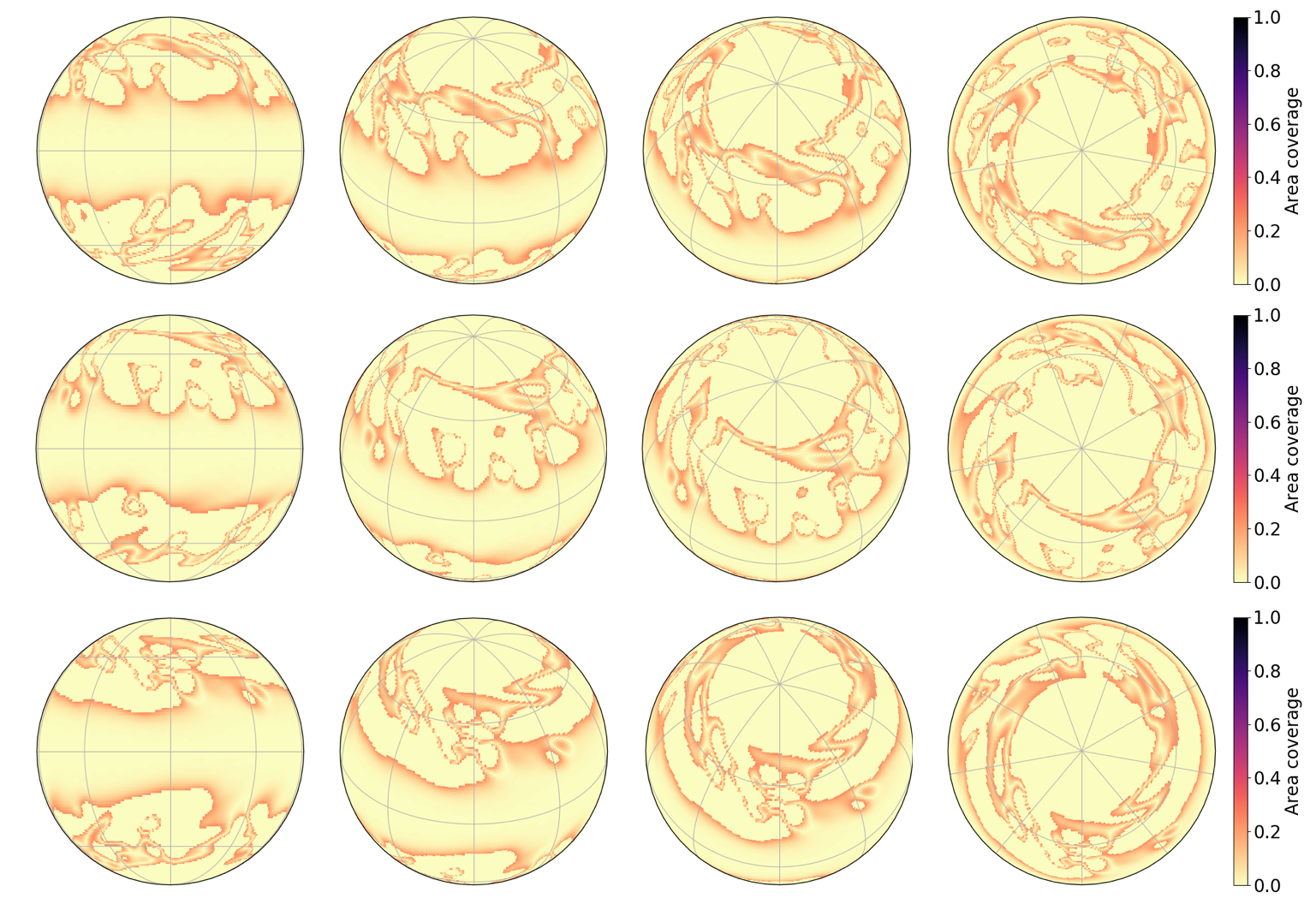}
\caption{Similar to Fig.~\ref{fig:map_spots}, but showing the faculae area coverages.}
\label{fig:map_faculae}
\end{figure}

We present maps for different inclinations and nesting degrees of the spot and facula distributions (following the description in Sect.~\ref{filling_factors} and the choice of the parameters presented just above) for a star with 8\rotrate{} as returned by FEAT with various nesting degrees (see Sect.~\ref{FEAT}) in Fig.~\ref{fig:map_spots} and Fig.~\ref{fig:map_faculae}, respectively. Clearly visible in Fig.~\ref{fig:map_spots} are the polar spots.

\section{Light curves}\label{lightcurves}

Using the model described in Sect.~\ref{FEAT}, we generated a synthetic 11--year--long cycle for each rotation rate considered in this work. 
We focus on the four years around the activity maximum for two reasons. Firstly, we aim to explain the upper envelope of the variability distribution of stars as a function of the rotation period \citep{McQuillan2014}. Secondly, we model a single activity cycle, without overlapping cycles preceding and following it. Modelling cycle overlap is beyond the present scope, and it would affect the activity level around cycle minima, which we thus exclude from the analysis. 
 
We note that the LCs shown in this section are 180--days--long snippets of the full LCs and were simply chosen for demonstration.

\begin{figure}[h!]
\centering
\includegraphics[width=\columnwidth]{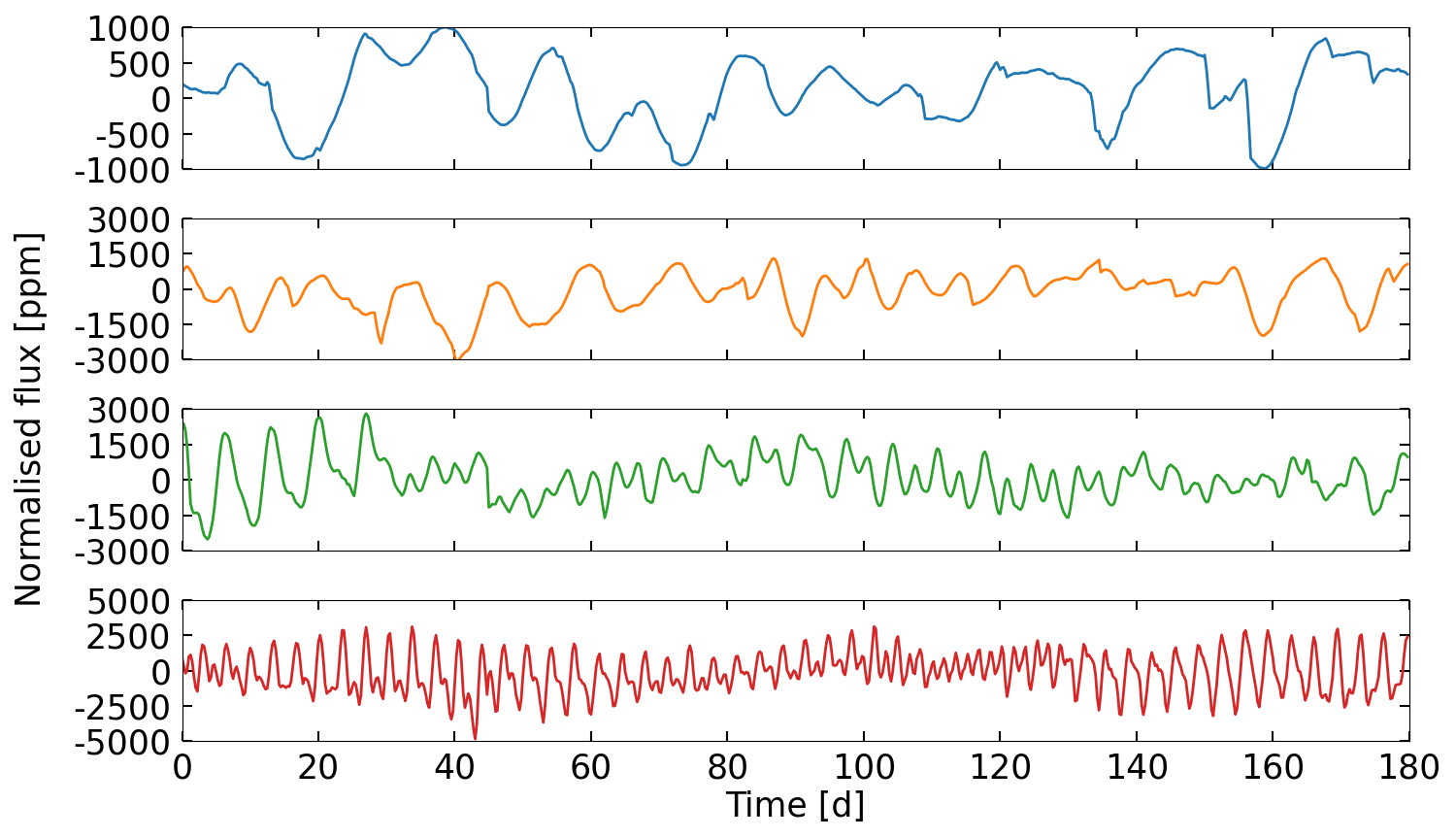}
\caption{Synthetic light curves (LCs) for stars with different rotation rates as they would be observed in the \textit{Kepler} passband at an inclination of $i=90\degree$. Shown are not-nested cases ($p=0$) with rotation rate values of 1\rotrate{} (blue),  2\rotrate{} (orange), 4\rotrate{} (green), and  8\rotrate{} (red).}
\label{fig:LCs_no_nest}
\end{figure}

Firstly, we consider the LCs of stars that are observed equator-on ($i=90^{\circ}$). Figure~\ref{fig:LCs_no_nest} displays the detrended LCs for the non-nested case ($p=0$), for different rotation rates. One can see that the faster the star rotates, the higher its amplitude of variability (mainly due to the shorter rotation periods - the lifetimes of the active regions change far less) which is a consequence of the activity-rotation scaling (see Step 2 from Sect.~\ref{FEAT}). The small bumps in the LCs that are mostly seen in the case for the
$  \Omega_{\star} = 1\Omega_{\odot} $ are a result of firstly having all BMRs emerge at their maximum size in the SFTM. Since the emergence rate is generally low and not many other BMRs are present, this affects the LCs more severely in this low activity case. Secondly, and lastly, due to the nature of the threshold approach outlined in Sect.~\ref{filling_factors} flux superposition (in case of same polarity encounters) and cancellation (in case of opposite polarity encounters) leads to the possibility of flux switching from being associated with faculae to spots (or spots to faculae) rapidly. Similar effects for the solar rotator will be visible in the following plots as well for the very same reason.

Figure~\ref{fig:LCs_no_nest} shows that not only amplitude but also the shape of the LCs strongly depends on the stellar rotation rate. For the case of the  solar rotation ($  \Omega_{\star} = 1\Omega_{\odot} $, top panel of Fig.~\ref{fig:LCs_no_nest}), most of the individual dips in the LCs correspond to transits of different active regions (since active regions evolve on timescale shorter than the solar rotation). As active regions emerge randomly in time, the LCs appear quite irregular. In contrast, the LCs of the more rapidly rotating stars show gradually more regular patterns in brightness variations (see the lower three panels in Fig.~\ref{fig:LCs_no_nest}). 
This is because active regions on such stars
can survive several rotation periods. Furthermore, the large amount of BMR emergences over mid- to high latitudes with large tilt angles leads to the formation of polar spots at about 4\rotrate{}, with prominent polar spot caps being present for the 8\rotrate{} rotator (see Fig.~\ref{fig:map_spots}). The formation of polar spots for stars with those rotation rates are consistent with \citetalias{Isik2018} and Doppler-imaging observations. We note that the polar spots in our simulations turn out to be non-axisymmetric unipolar caps. Their overall structure is rather stable, because their decay is compensated by the magnetic flux coming from the new emergences, as long as the activity level and the BMR polarity orientations are sustained.
As a next step, we consider models with active-region nesting. 

\begin{figure}[h!]
\centering
\includegraphics[width=\columnwidth]{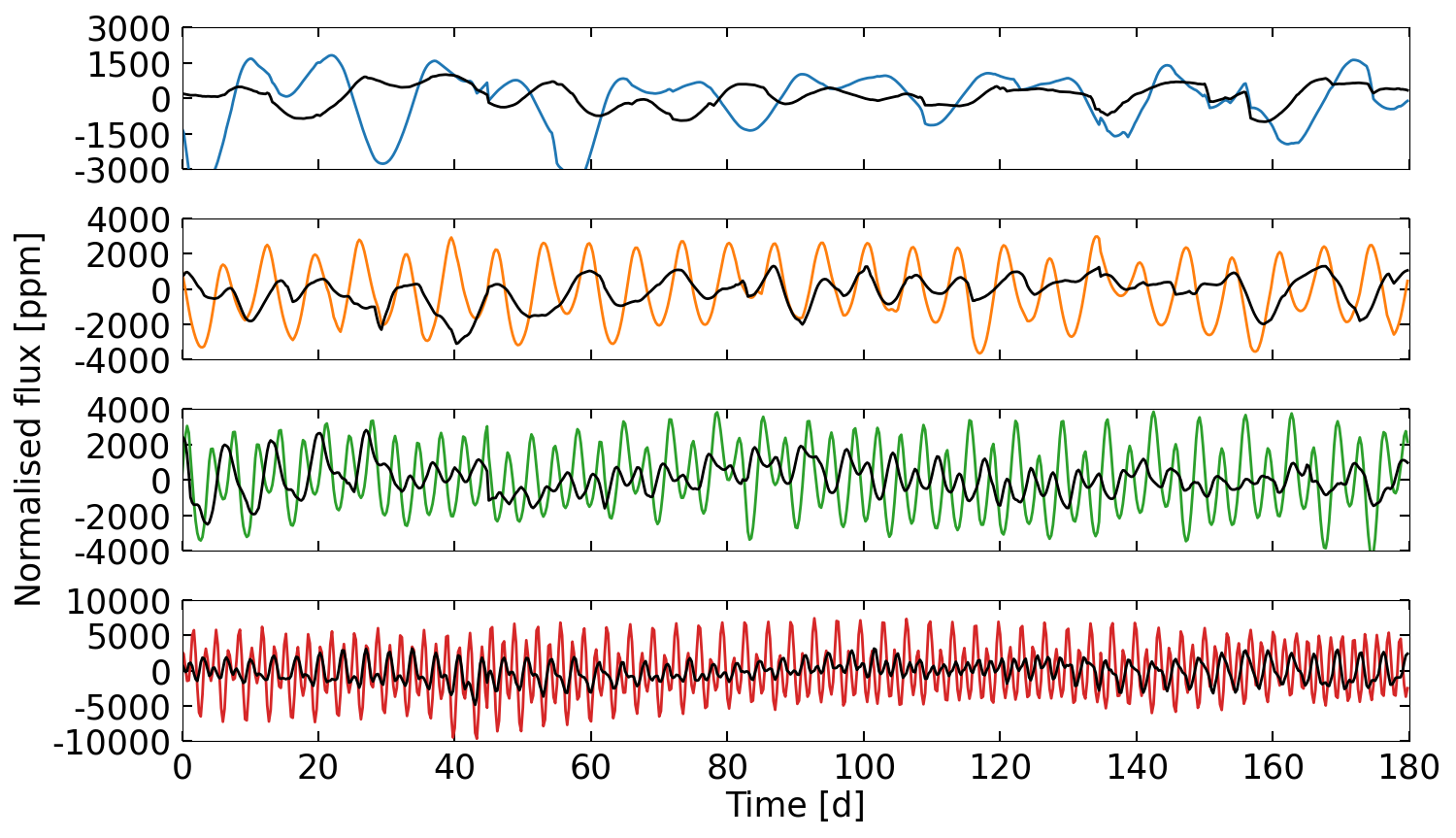}
\caption{Similar to Fig.~\ref{fig:LCs_no_nest}, where the black curves represent the calculations with $p=0$ for each rotation rate and the coloured curves represent those with added AL-type nesting at $p=0.7$.}
\label{fig:LCs_AL_70}
\end{figure}

We first consider the effect of the active-longitude nesting (see Sect.~\ref{FEAT}). Figure~\ref{fig:LCs_AL_70} shows LCs synthesised with AL nesting of 70\% (i.e.  $p=0.7$).
The overall shape of the LC for 1\rotrate{} is still rather irregular, compared to the faster rotators, due to the low emergence frequency. With increasing rotation rate, dips related to BMR transits not only occur at a separation of the rotation period, but also at half of the rotation period. In addition to a change in the morphology of the LCs, the variations are amplified with respect to the corresponding non-nested case at each rotation rate (black solid lines versus coloured lines).  

\begin{figure}[h!]
\centering
\includegraphics[width=\columnwidth]{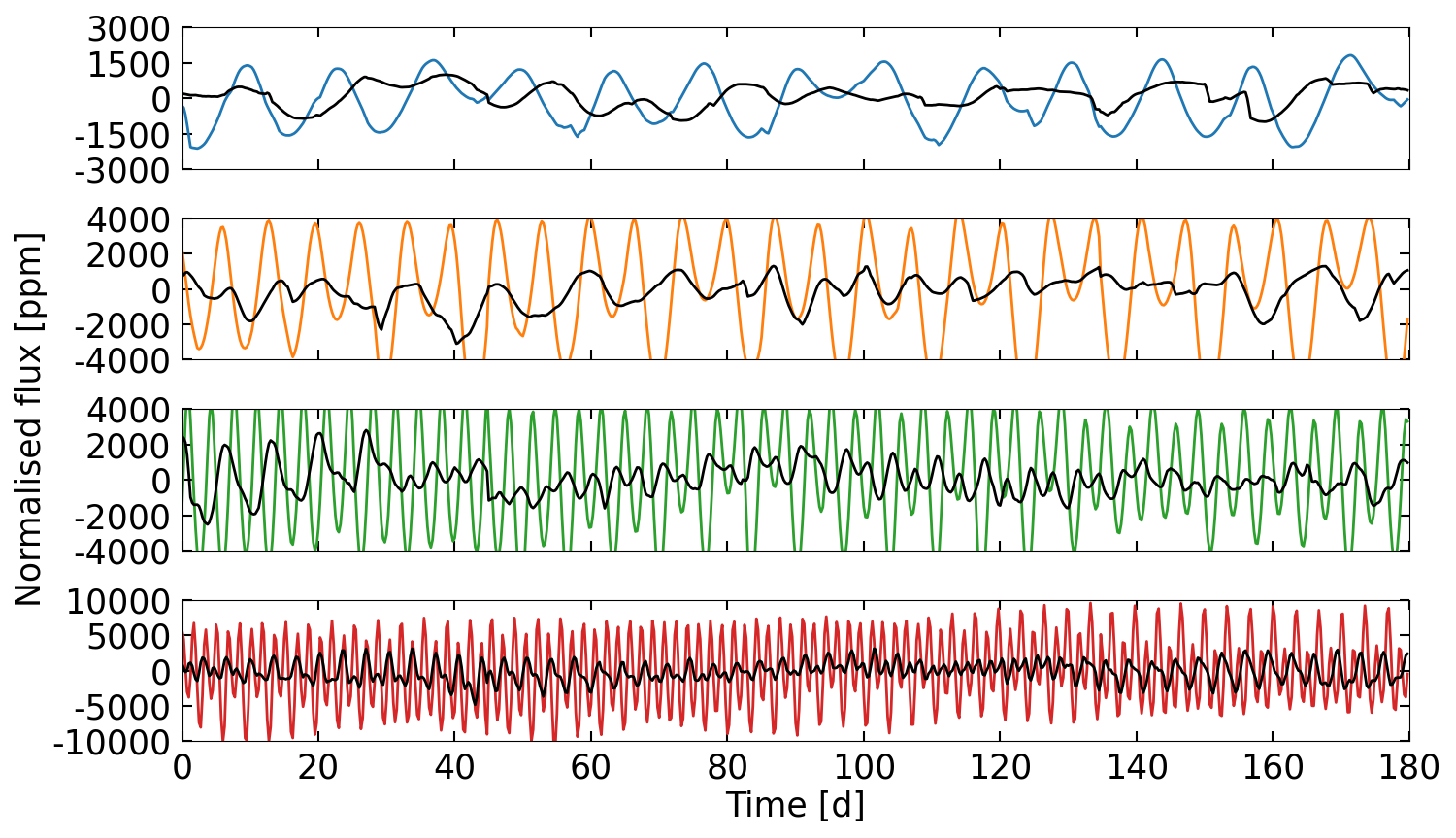}
\caption{Similar to Fig.~\ref{fig:LCs_AL_70} with AL-type nesting and $p=1$.}
\label{fig:LCs_AL_100}
\end{figure}

Figure~\ref{fig:LCs_AL_100} gives the most extreme case of AL nesting, where we assume that all BMRs emerge in one of the two active longitudes (e.g. $p=$1).  Clearly, both dips (at one and half-rotational period interval) occur
in all of the cases and the LC amplitudes are further augmented. The LCs in Fig.~\ref{fig:LCs_AL_100} additionally show that the two peaks have almost the same amplitude for the faster rotators. In case of the solar rotator and the low emergence frequency, temporarily asymmetries between the two AL can arise in case of the emergence of a large BMR. However, with increasing emergence frequency, the two ALs become more and more symmetric and hence the amplitude of the two dips due to the ALs as the star rotates are, to a large extent, comparable.  

\begin{figure}[h!]
\centering
\includegraphics[width=\columnwidth]{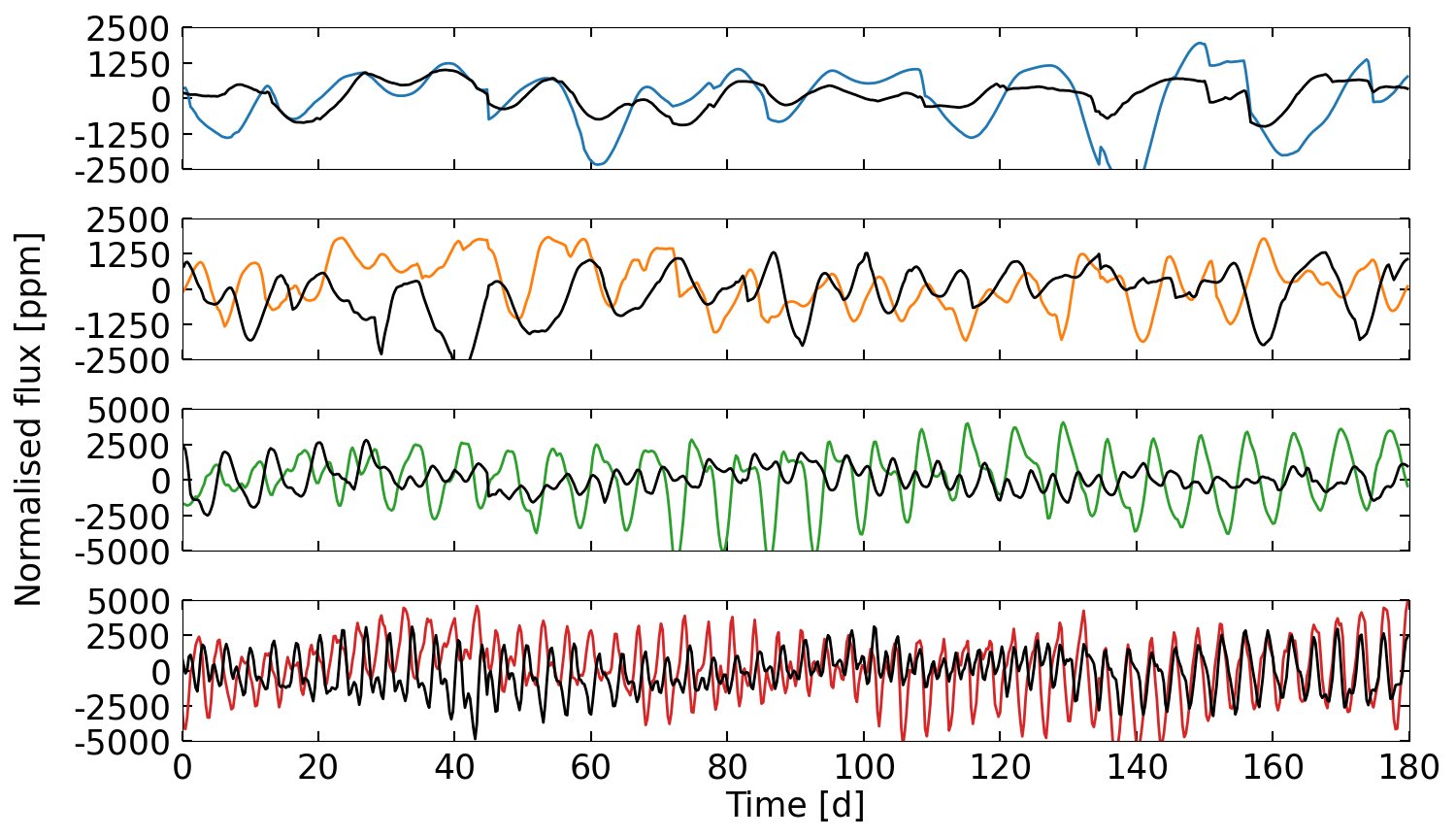}
\caption{Similar to Fig.~\ref{fig:LCs_AL_70}, with free nesting and $p=0.7$.}
\label{fig:LCs_FN_70}
\end{figure}

\begin{figure}[h!]
\centering
\includegraphics[width=\columnwidth]{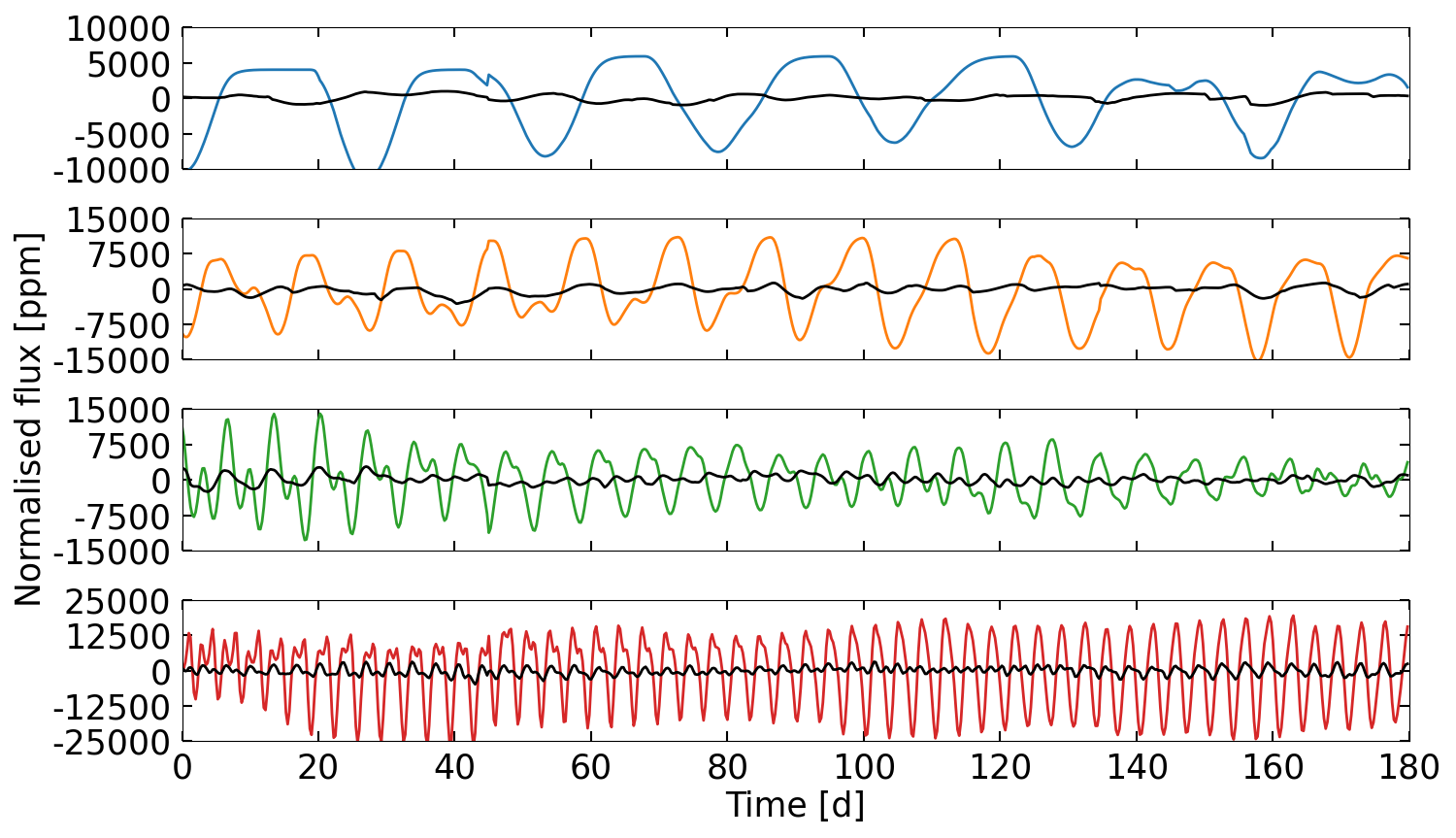}
\caption{Similar to Fig.~\ref{fig:LCs_AL_70}, with free nesting and $p=0.99$.}
\label{fig:LCs_FN_99}
\end{figure}

We now focus on the  behaviour of the LCs in the FN mode.  In Fig.~\ref{fig:LCs_FN_70} we present the LCs produced with $p=0.7$. One can see that the amplitudes of the variability increase for all four shown rotation periods. The light curves also appear more regular than those calculated with $p=0$.
We increase the nesting even further to $p=0.99$ in Fig.\ref{fig:LCs_FN_99}. The change in the LCs with respect to the non-nested case is remarkable. The amplitude of the LCs is enhanced for all cases and the runs with $p=0.99$ exhibit regular patterns, even in the solar case (top panel of Fig.~\ref{fig:LCs_FN_70}). Interestingly, in the displayed LCs, not only dips with separation of the rotation period, but also half-of the rotation period (e.g. most prominently seen in the 4\rotrate{} star) appear. This is interesting, as the dips at half of the rotation period, appear and disappear and are not constant, compared to the AL case. We will discuss this further in Section 5.

\begin{figure}
\centering
\includegraphics[width=\columnwidth]{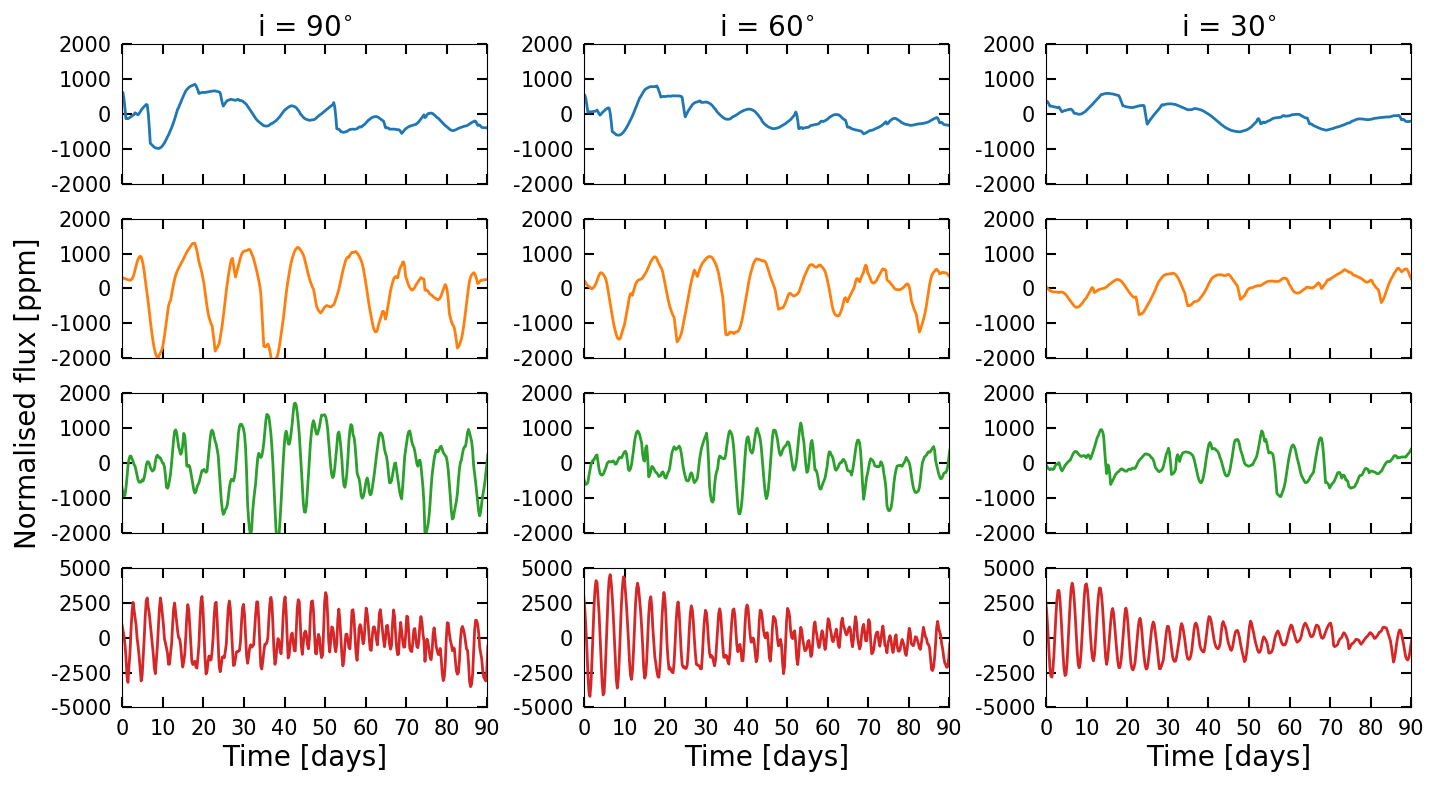}
\caption{Similar to Fig.~\ref{fig:LCs_no_nest} for different inclination shown in three columns for $i=90\degree$, $60\degree$, and $i=30\degree$.}
\label{fig:LC_incl}
\end{figure}

Next, we consider the inclination effect on the LCs. For demonstration, we limit ourselves to the non-nested cases with different rotation rates. 
In Fig.~\ref{fig:LC_incl}
we show the time-span of 0--90 days from Fig.~\ref{fig:LCs_no_nest} for inclinations of $90\degree$, $60\degree$ and $30\degree$. In the given timespan, for 1\rotrate{}, the amplitude of the variability decreases with decreasing inclination. The shape of the transits also change. 
For 2 and 4\rotrate{}, the LC amplitudes decreases for the inclinations shown here as well. Interestingly, the situation changes for the  8\rotrate{} case. The amplitude increases from $i=90\degree$ to $60\degree$ and then decreases from $i=60\degree$ to $i=30\degree$. Also the amplitude of variability observed at  $i=30\degree$ is larger than that at $i=90\degree$. These inclination dependencies can  be explained with the help of the magnetic field maps in Fig.~\ref{fig:map_spots}. For the case of 1\rotrate{} all regions emerge within $\pm$ 30\degree around the equator. The Coriolis effect gets stronger with increasing rotation rate, so that for the 8\rotrate{} star, the BMRs can emerge at latitudes up  to $\pm$ 70\degree, while a latitudinal belt free of active regions opens up around the equator between $\pm 20^\circ$ latitudes (see Sec.~\ref{discussion}). For $i=90\degree$, the high-latitude spots appear close to the limb, where their effect on the brightness is significantly reduced due to the foreshortening. 
With decreasing inclination, the majority of spots shift towards the centre of the visual disc and their effect on brightness increases, so that at intermediate inclinations, the variability reaches a maximum (see Figs.~\ref{fig:comp_timo_1} panel a), before it starts decreasing again, as the spots move towards the limb of the visible disc once more.

\section{Comparison to observations}\label{obs}

In the following, we compare our model to observational records of Sun-like stars obtained by the \textit{Kepler} telescope. Since our model builds on the solar paradigm, we select stars with near-solar effective temperatures between 5500--6000\,K and surface gravity $log\,g \geq 4.2$ using the updated fundamental parameters of \textit{Kepler} stars from \citet{Berger2020}. Next, we consider stars with known rotational periods using the data sets of \cite{McQuillan2014} (hereafter \citetalias{McQuillan2014}) and \cite{Santos2021} (hereafter \citetalias{Santos2021}). These constraints lead to a sample of 6,228 stars when using \citetalias{McQuillan2014} rotation periods and 11,493 stars when using \citetalias{Santos2021} rotation periods.

We express the variability through the quantity $R_{var}$, first introduced by \cite{Basri2010,Basri2011}. We use a slightly modified version of the range $R_{var}$: we compute the difference between the 95th and 5th percentile of the sorted differential intensities. Even for the latest \textit{Kepler} data reduction (DR25) it occurs that some quarters still include instrumental systematics that might influence the range. We then compute the median absolute deviation (MAD) of all $R_{var}$ values, and remove those quarters, that deviate more than 6x the MAD. Afterwards the median is taken of all remaining quarters, which we call $R_{var}$. Since the SFTM returns instantaneous values with 6-hour cadences, we take every 12th data point (the \textit{Kepler} long cadence is $\approx$30 min) and compute $R_{var}$ for this unbinned time series. All computations are based on the latest \textit{Kepler} data release (DR25) using the PDC-MAP light curves. We also cross-checked our calculated variabilities represented through $R_{var}$, with the metric $R_{per}$ used by \cite{McQuillan2014} and found very good agreement.

Before computing $R_{var}$ of the models, we need to add noise to the light curves. We follow the same  strategy as described in detail in \citet{Timo2020}. Here, we multiply the noise with a factor of $\sqrt{3h/30min}=\sqrt{6}$ to account for the different time bins. For each inclination, 1000 noise realizations are considered, and the mean and standard deviations are computed. This has been done only for the 1\rotrate{} $p=0$ case, as we found that the noise level is significantly lower than the actual stellar variability in all other cases.

\begin{figure}[!h]
\centering
\includegraphics[width=.48\columnwidth]{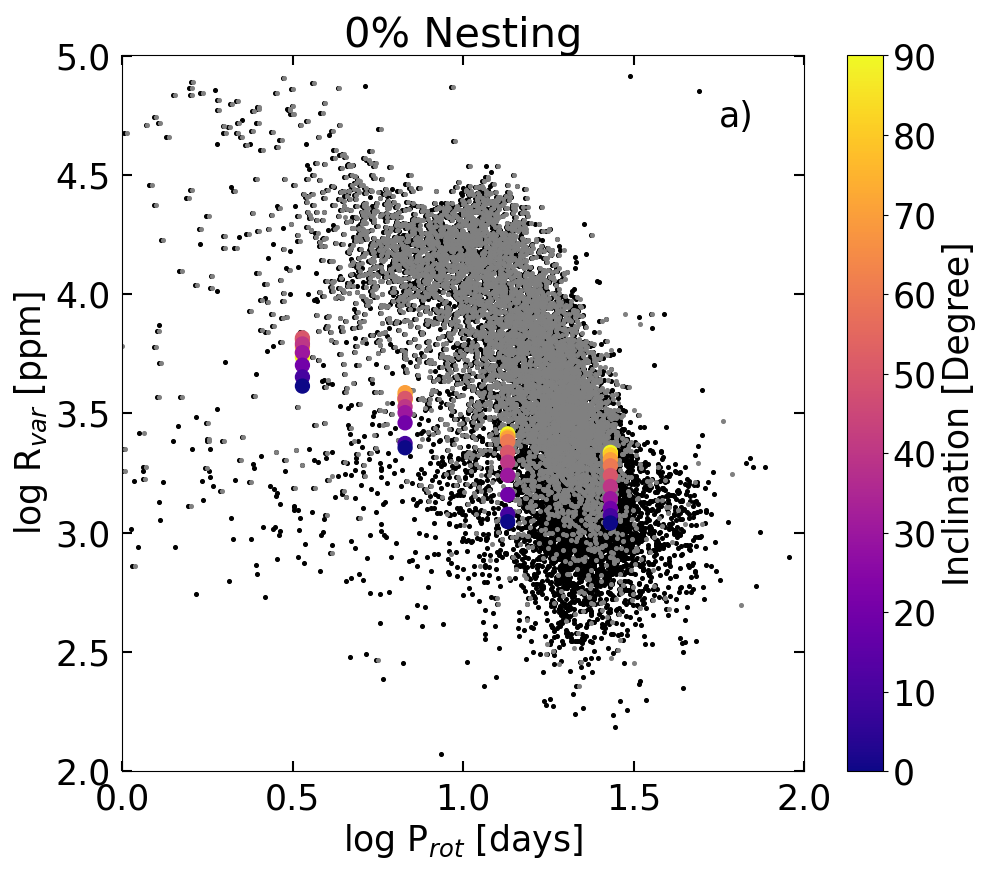}\quad
\includegraphics[width=.48\columnwidth]{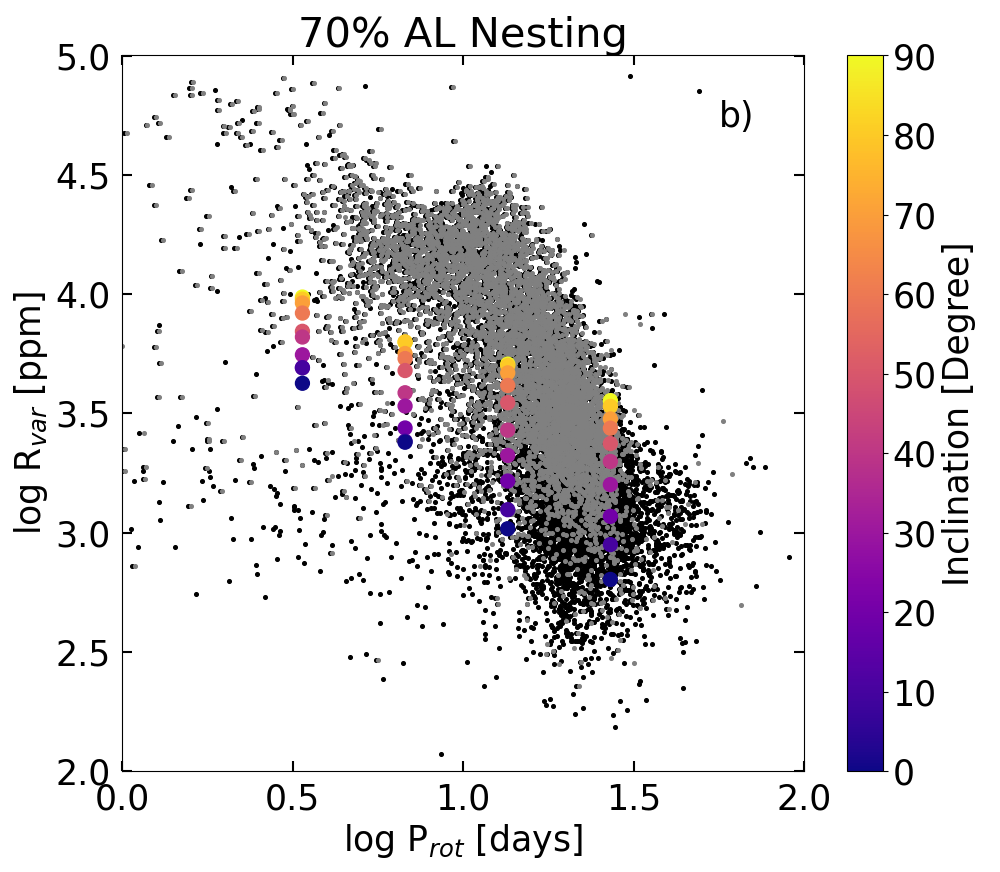}
\medskip
\includegraphics[width=.48\columnwidth]{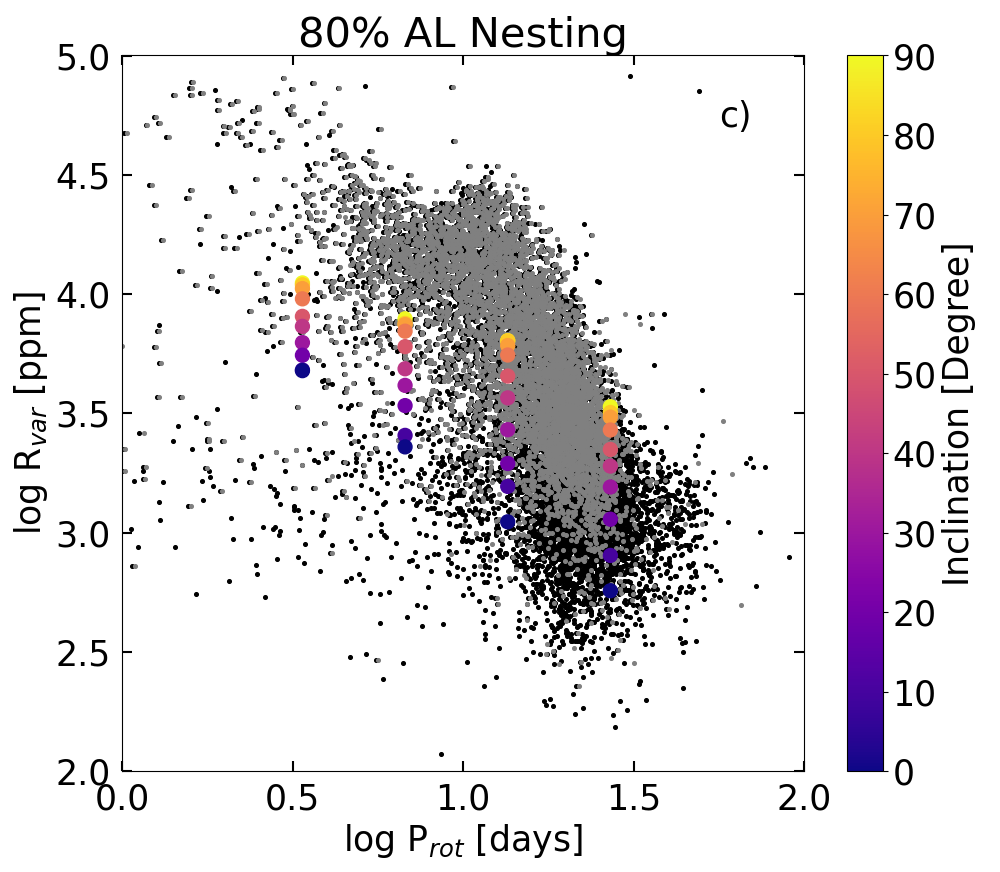}\quad
\includegraphics[width=.48\columnwidth]{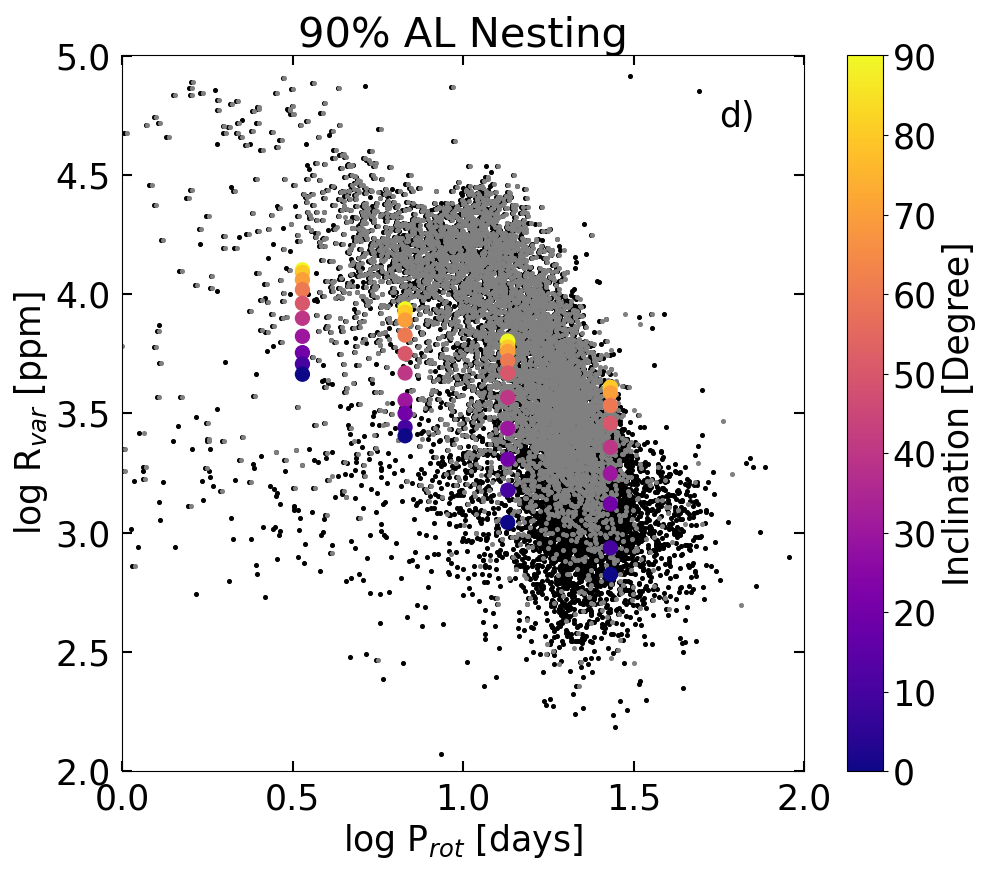}
\medskip
\includegraphics[width=.48\columnwidth]{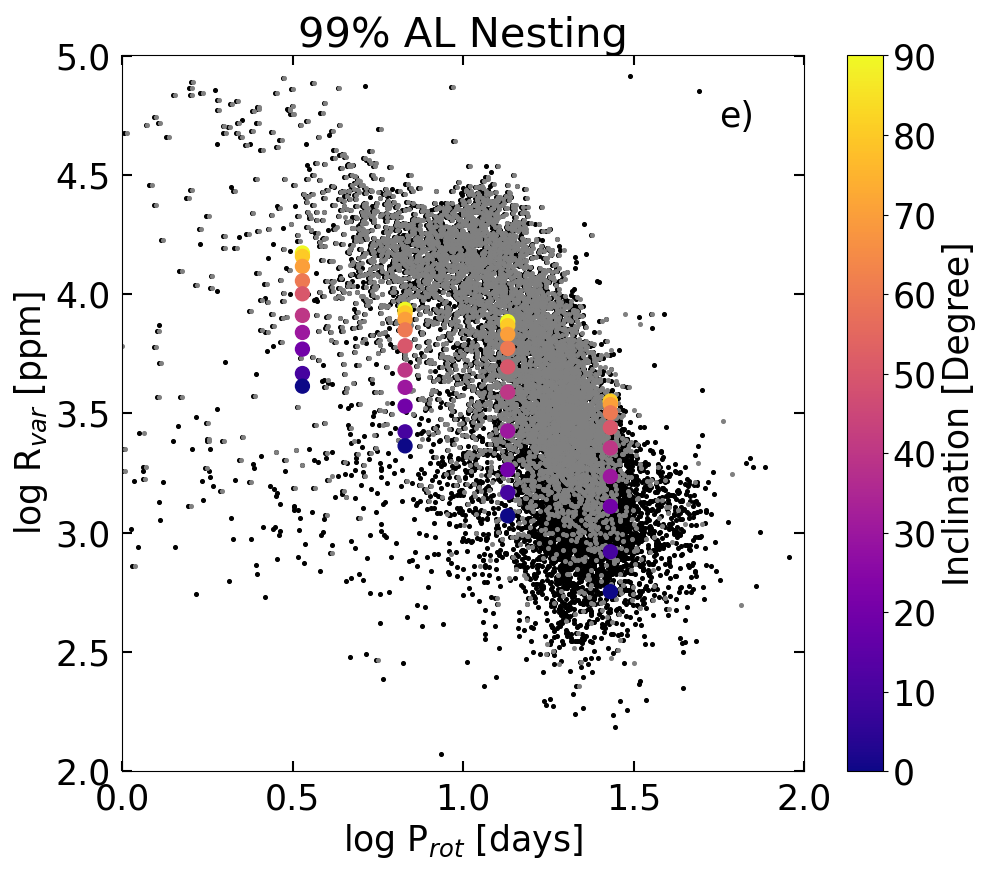}
\caption{Comparison of \rvar{} as a function of the rotation period between stars with effective temperatures with 5500--6000 K and \textit{log g $>$} 4.2 with detection rotation periods from \citetalias{McQuillan2014} (grey dots) and \citetalias{Santos2021}  (black dots) and the modelled stars. Each panel includes different nesting probabilities $p$ in the form of active-longitude (AL) nesting. The different colours indicate the inclination of the modelled stars.
\label{fig:comp_timo_1}}
\end{figure}

\begin{figure}[!h]
\centering
\includegraphics[width=.48\columnwidth]{distr_0_nest_both.png}\quad
\includegraphics[width=.48\columnwidth]{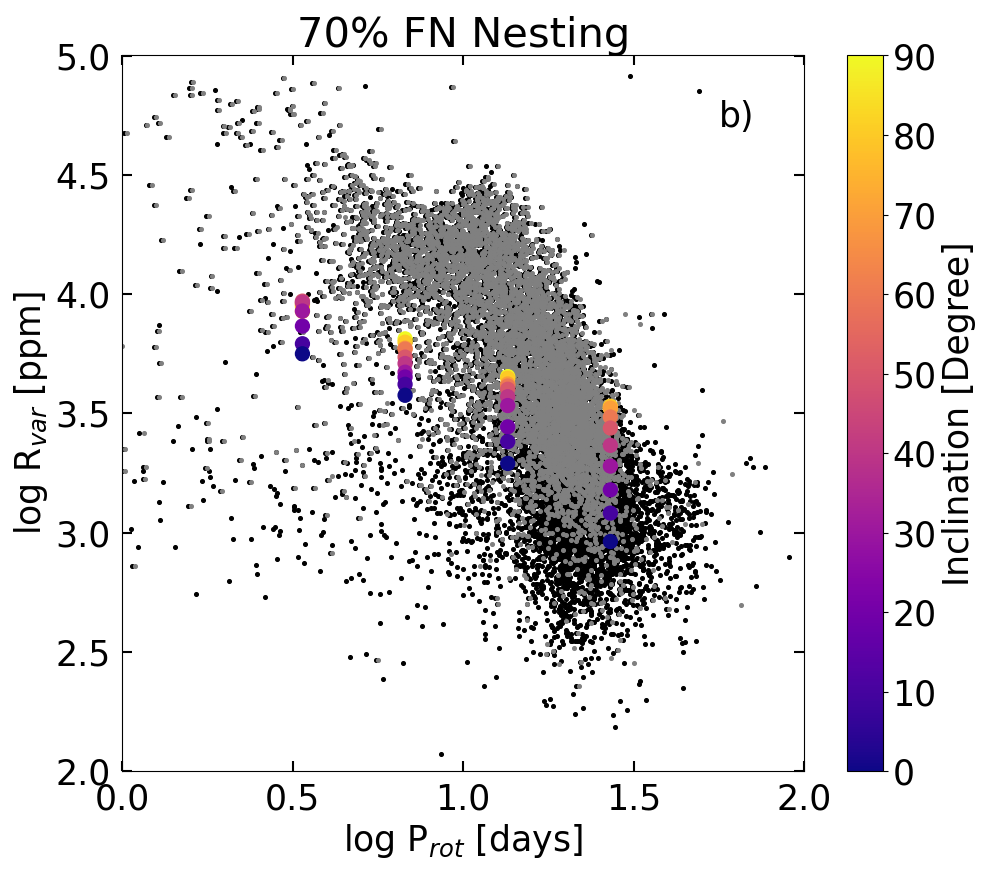}
\medskip
\includegraphics[width=.48\columnwidth]{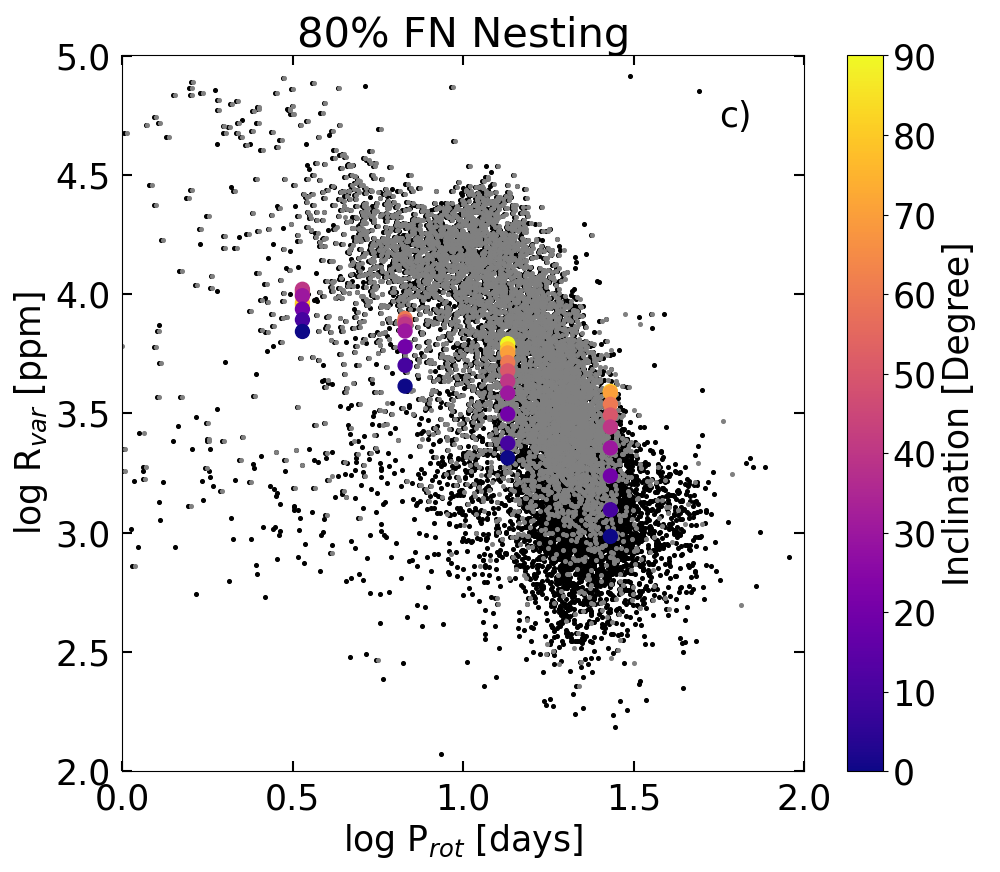}\quad
\includegraphics[width=.48\columnwidth]{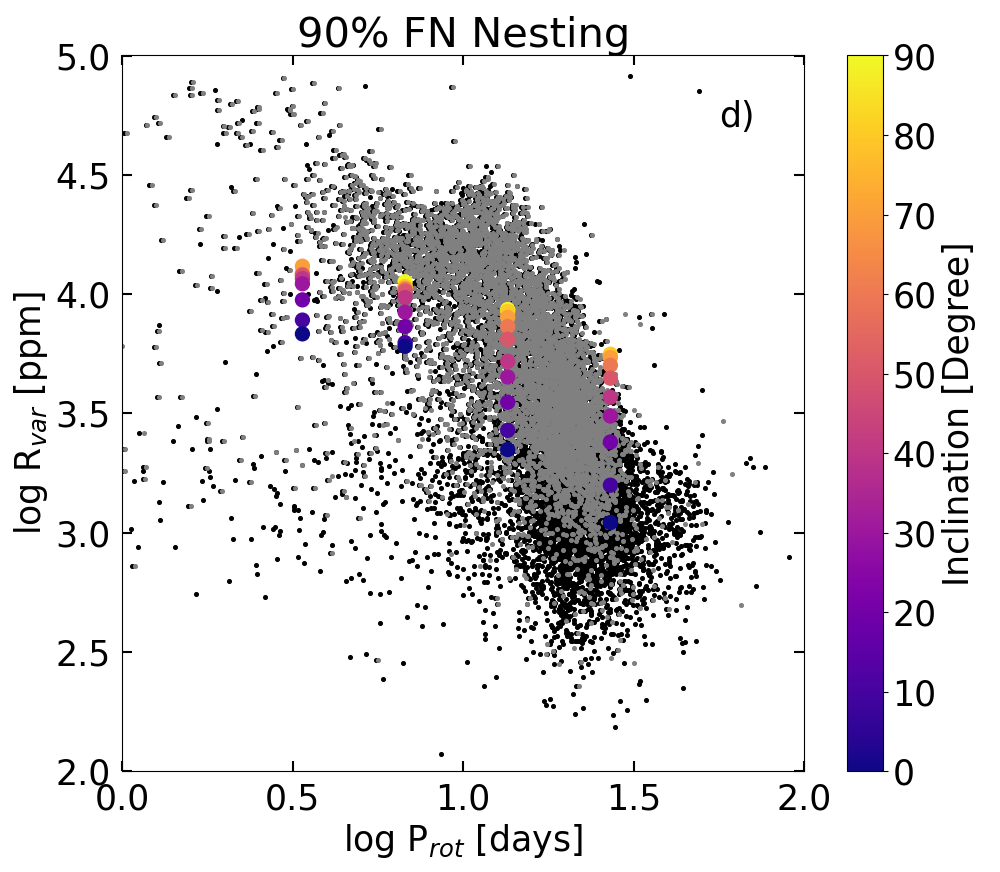}
\medskip
\includegraphics[width=.48\columnwidth]{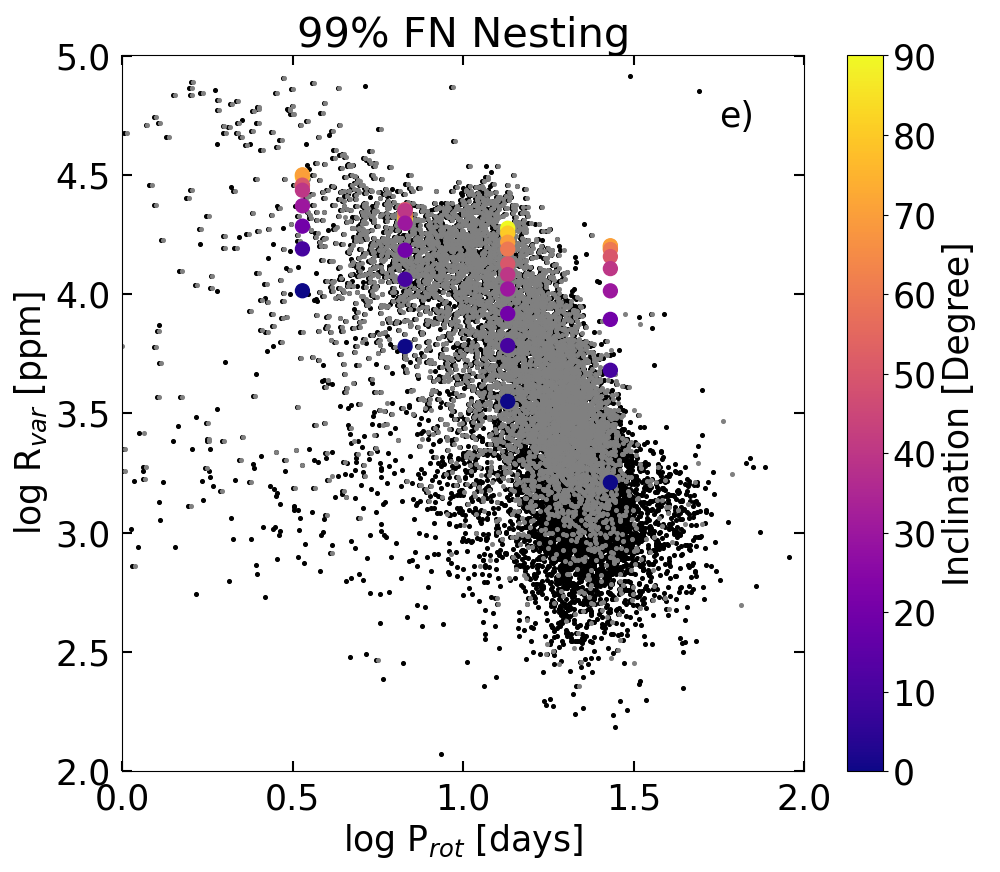}
\caption{Similar to Fig.~\ref{fig:comp_timo_1}, with different degrees of free nesting (FN).}
\label{fig:comp_timo_2}
\end{figure}

In Figs.~\ref{fig:comp_timo_1}~and~\ref{fig:comp_timo_2} we show the comparison between the calculated \rvar{} values of our simulated stars for various degrees of nesting in the AL and the FN mode, respectively, and the two samples of \textit{Kepler} stars shown as grey (\citetalias{McQuillan2014}) and black (\citetalias{Santos2021}) dots. Evidently, if we do not include any nesting ($p=0$), our calculated variabilities underestimate the bulk of the observed variability amplitudes, especially for the faster rotators. With increasing nesting level (Fig.~\ref{fig:comp_timo_1} and \ref{fig:comp_timo_2} b, c, d, and e), \rvar{} increases and the values move towards the upper edge of the distribution. Interestingly, $p=0.99$ in the FN mode (Fig.~\ref{fig:comp_timo_2}~e) overestimates the variability of the solar rotators but leads to similar variability values as the upper envelope of the variability distributions of the faster rotating stars. We note that, while the numbers of BMR emerging is the same for a given rotation rate between the non-nested and nested cases, with higher nesting degree, the spot disc area coverage increases due to the formation of spots by flux superposition. As a consequence, the spot area coverage is not preserved in contrast to the approach presented in \cite{Isik2020} and the nesting has a stronger effect on variability in our model than in \cite{Isik2020}. We will elaborate on this point further in Sect.~\ref{discussion}.

\begin{figure*}[!h]
\centering
\includegraphics[width=.23\linewidth]{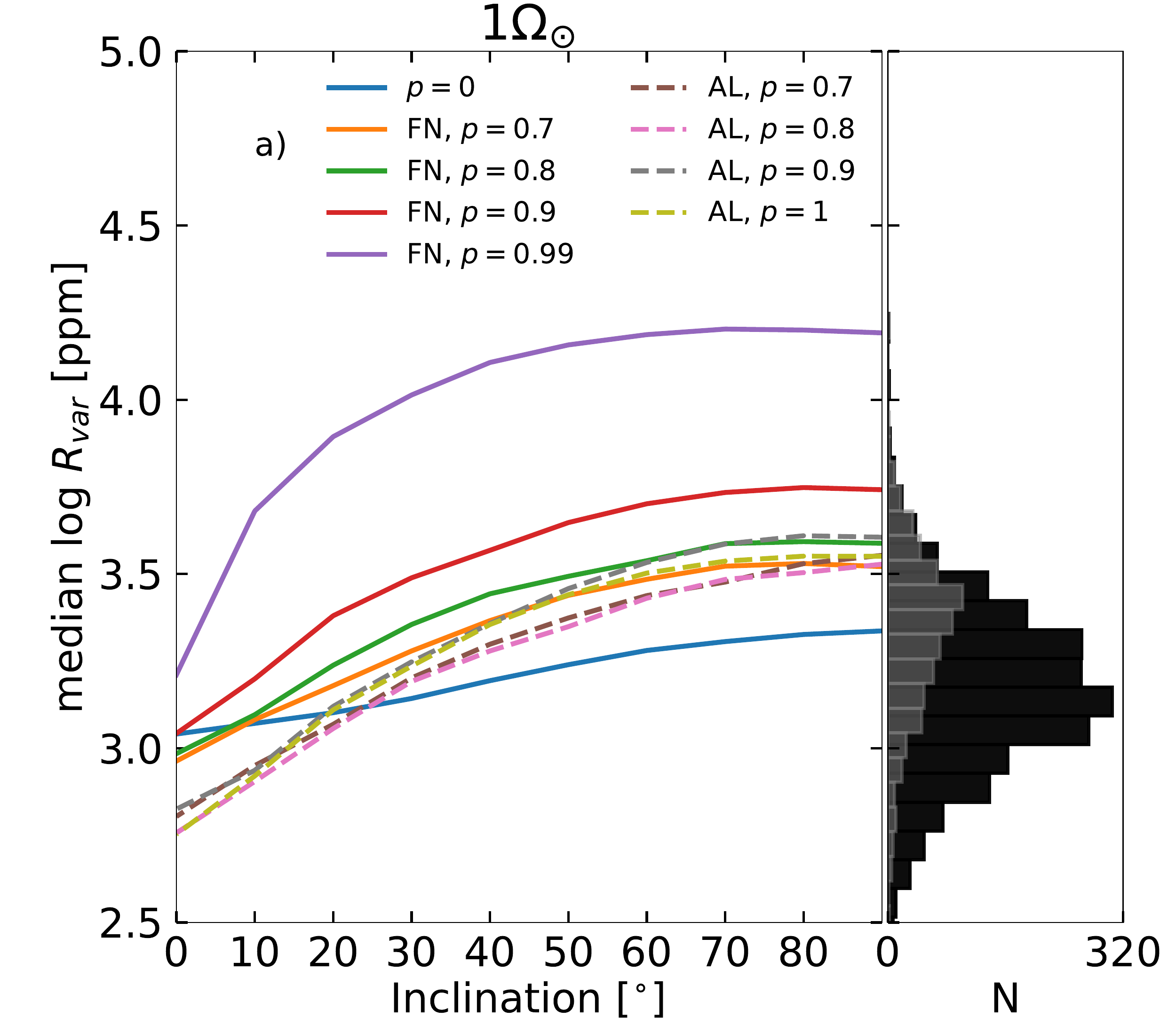}
\includegraphics[width=.23\linewidth]{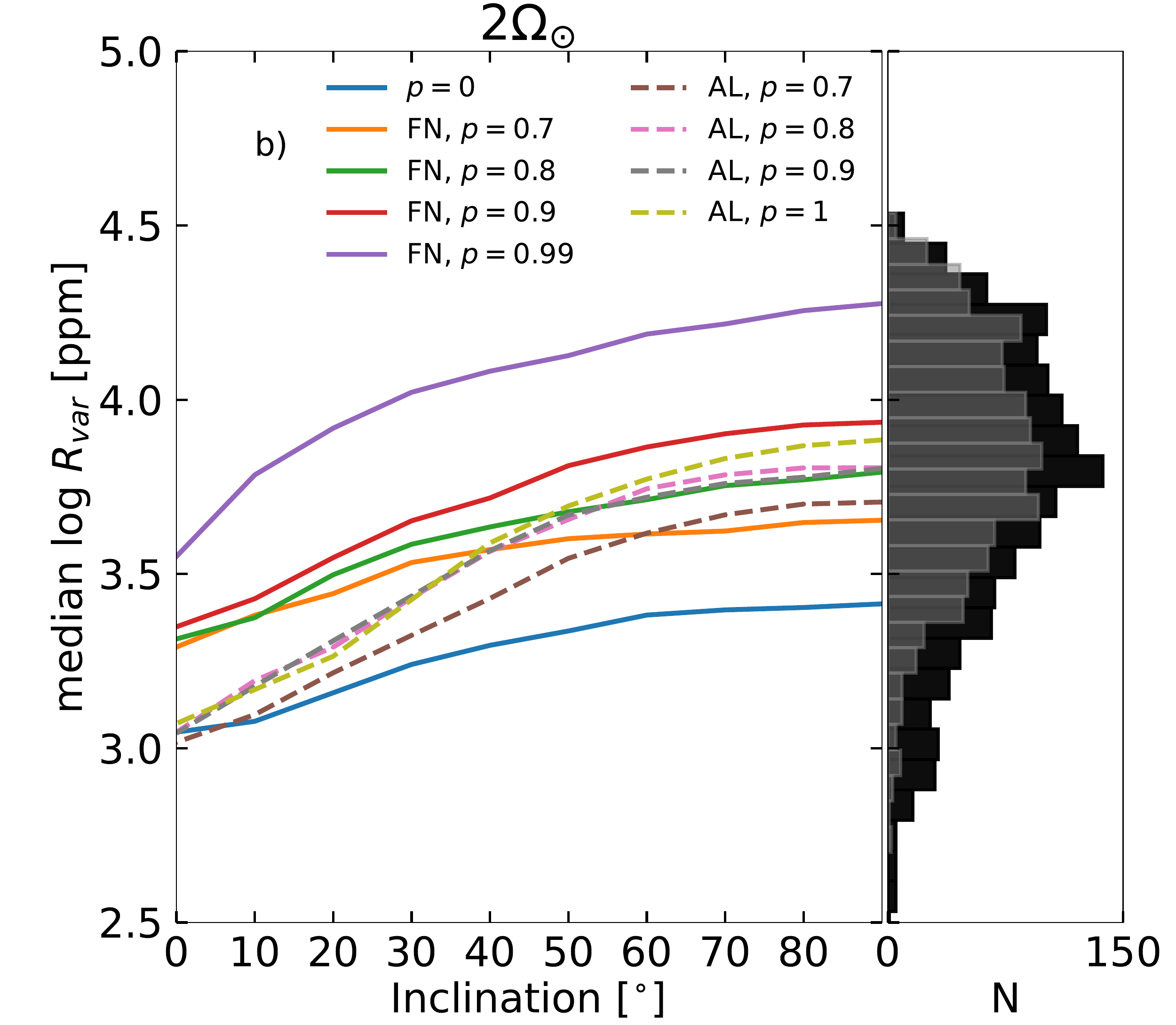}
\includegraphics[width=.23\linewidth]{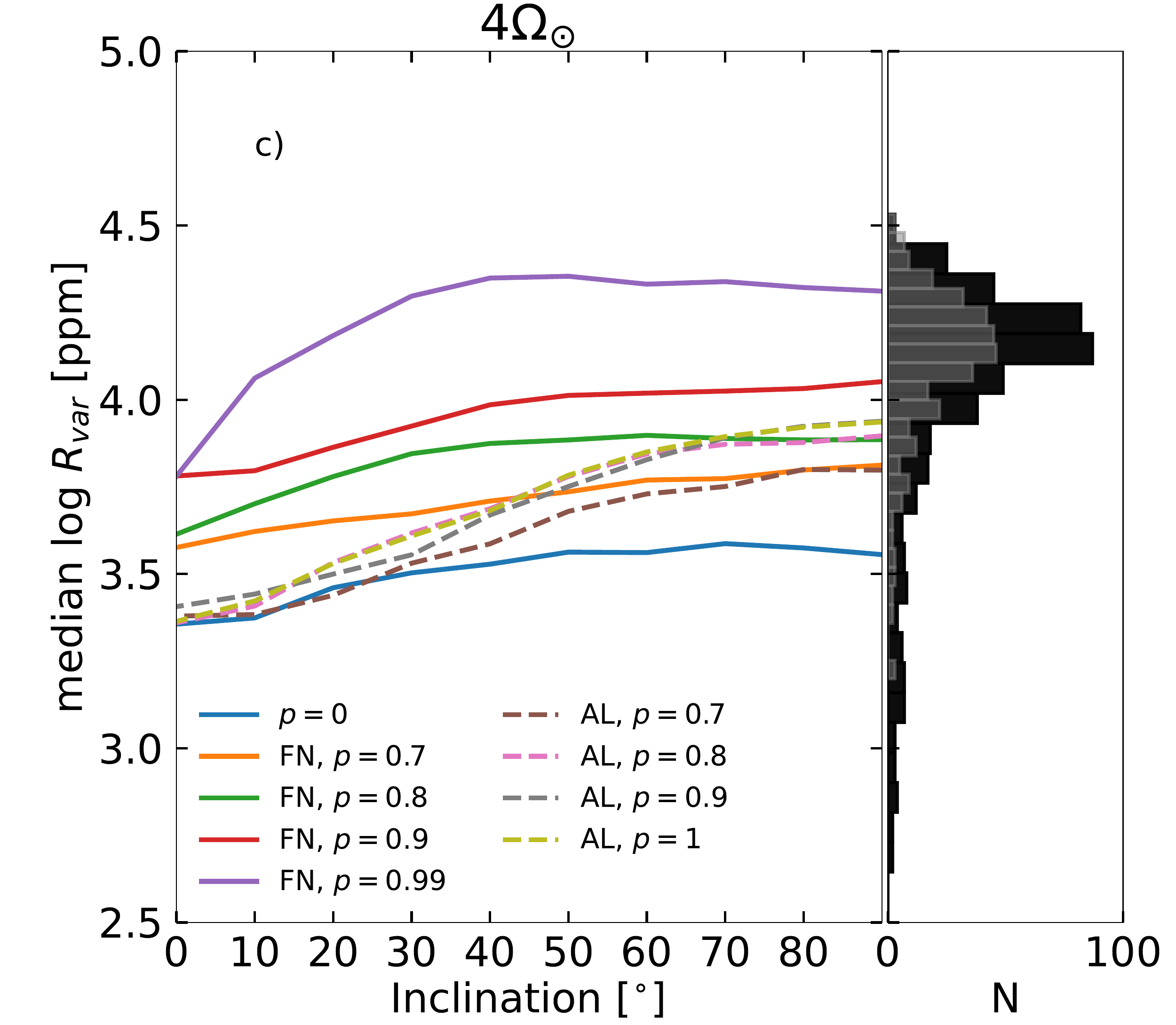}
\includegraphics[width=.23\linewidth]{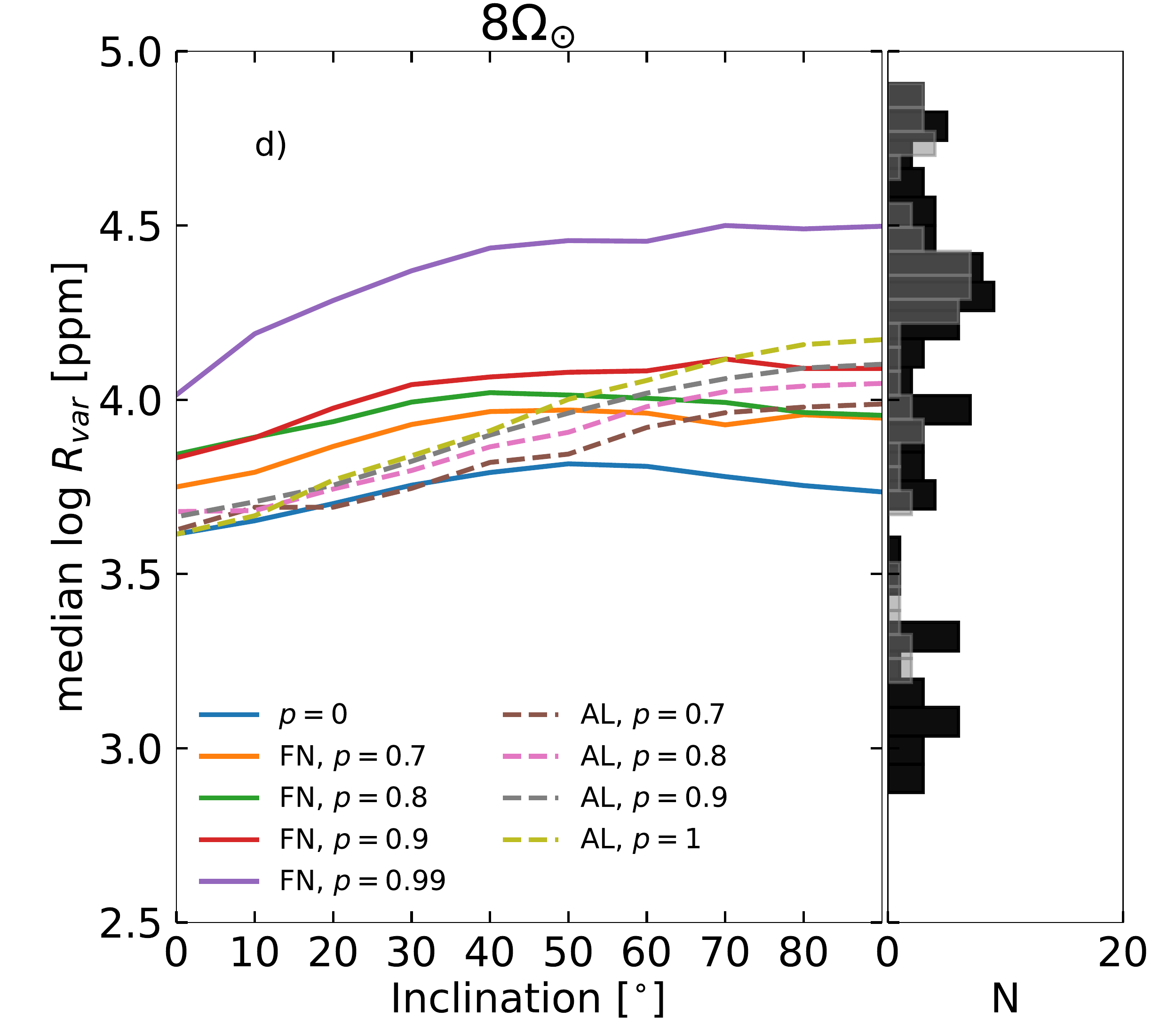}
\caption{Dependence of \rvar{} on the inclination. Each panel represents different realisations of the BMR emergence for a given rotation rate. The grey histograms are drawn from the \citetalias{McQuillan2014} stars and the black histograms from the \citetalias{Santos2021} stars, with the limitations of [23/X, 27/X] days for a given rotation rate X in units of the solar rotation (i.e. for X$=$2, meaning stars with 2\rotrate{}, the stars have rotation periods between [11.5,13.5] days).}
\label{fig:histograms}
\end{figure*}

For a more detailed comparison between the simulations and the observations, we bin the distribution of observed variabilities. Namely, we compare the variabilities returned by our model for a star rotating X times faster than the Sun, with variabilities of a sample of stars with periods in the range [23/X, 27/X] days from the \textit{Kepler} samples. This comparison is shown in Fig.~\ref{fig:histograms}. The histograms in grey and black display the range of variabilities within each of the rotation period bins. We note that the number of stars within each rotation bins decreases from panel a to d. Similarly to Figs.~\ref{fig:comp_timo_1} and \ref{fig:comp_timo_2}, Fig.~\ref{fig:histograms} shows that the calculations with $p=0$ clearly underestimate the variability for the 2\rotrate{} and 4\rotrate{} sub-samples (while the small number of {\it Kepler} stars in the 8\rotrate{} rotation bin makes interpretation of panel d rather difficult).

For the stars with near-solar rotation periods (i.e the 1\rotrate{} case), there is a substantial difference between the stars in \citetalias{McQuillan2014} and \citetalias{Santos2021}. The sample of \citetalias{Santos2021} contains many more stars with variabilities lower than the solar variability at the maximum of cycle 22 (blue curve in the left panel of Fig.~\ref{fig:histograms}), while both samples roughly contain the same number of stars substantially more variable than the Sun (i.e. the high-variability tail). This result again raises the question of whether the Sun could also become as variable as those stars in the high-variability tail \citep{Timo2020}. Moreover, it agrees with the conclusions drawn in \cite{Timo2020, Timo2020_2} that the solar variability is not unexpectedly low but that the rotation periods of many stars with similar variabilities have just been missed by previous rotation period surveys (such as \citetalias{McQuillan2014}).

At the same time, the $p=0.99$ FN calculations in Fig.~\ref{fig:histograms} lie towards the upper bound of the variability distribution of the faster rotators, while highly overestimating the variability of stars with near-solar rotation periods. One can see that different nesting modes can lead to similar values of median \rvar{}, especially if the inclination of the stellar rotation axis is not known. This degeneracy might be lifted if alternative metrics are used. We discuss further metrics of characterising stellar variability including the morphology of the LCs in Sect.~\ref{discussion}. 


\section{Discussion}\label{discussion}

The premise of this work was to extend the solar paradigm to model the distribution of magnetic features on stars rotating faster than the Sun to use these to consecutively calculate the stellar light curves and the amplitude of the variability. Figure~\ref{fig:histograms} shows that while our calculations with a rather high degree of nesting can reproduce the bulk of variabilities in the {\it Kepler} sample they do not catch the maximum of the variability distribution. This might be because we have used the emergence frequency of BMRs for solar cycle 22 as reference cycle and scaled the emergence frequency as a function of the rotation period based on that cycle (i.e. a star rotating twice as fast as the Sun exhibits twice as many BMR emergences). First, solar cycle 22 does not represent the maximum level of activity the Sun is capable of \citep[see, e.g.][]{Usoskin_LR}. Second, the activity-rotation scaling itself is rather approximate \citep[see][for more details]{Isik2018}. 
We refer here to \cite{Isik2020}, who considered the  effect of activity increase on the variability with a considerably simpler model limited to stars with near-solar rotation rates. 

Empirical studies suggest that faster rotating stars have activity cycles shorter than the present Sun \citep{Boehm-Vitense2007}. In spite of this, as the main purpose of this study is to model stellar variability on the rotational timescale, we follow the approach of \citetalias{Isik2018} and assume that the solar temporal profile of activity remains unchanged for the faster rotators, for simplicity. However, we only consider the maximum level of activity (in a four-year window during activity maximum, see Fig. \ref{fig:butterfly_rots}). Thus, our light curves (and the resulting peak-to-peak variability) correspond to the maximum of the stellar activity cycle and are largely unaffected by the temporal profile of the underlying activity cycle. 

Another parameter space that we have not taken into account in this study is the effect of stellar fundamental parameters (e.g. the effective temperature and the stellar metallicity). While we limited our sample of \K{} stars to a temperature range of 5500--6000K, it contains stars with a rather broad range of metallicity values. \cite{Witzke2020} have shown that changing the metallicity for a star with solar level of activity enhances the possibility of recovering its rotation period, as the star moves out of the compensation regime between facular and spot contribution on the rotational timescale. It is less intuitive, however, how the change in metallicity will affect the calculations presented in this paper. According to our calculations, the rotational variability in the rapid rotators is primarily driven by spots, yet the spot component is less affected by the change in the metallicity than the facular component \citep[see][for a more coherent discussion]{Witzke2018}. However, metallicity might have an impact on the activity of a star and on the surface distribution of magnetic features \citep[][]{Amardetal2018, Amardetal2020, Victoretal2021} - an effect that we deem to be outside the scope of the present study.

The latitude of emergence calculated by flux-tube simulations (see Sect.~\ref{model} and \citetalias{Isik2018}) puts a well-defined lower limit for the latitude of emergence, which becomes more visible for fast rotators (see Fig.~\ref{fig:butterfly_rots}). This is a consequence of the inward-directed Coriolis force in the rotating frame, consistently acting on rising flux tubes having a pro-grade azimuthal flow (i.e., they rotate faster than their locality). This leads to a well-defined minimum latitude of emergence, corresponding to the minimum non-zero latitude of injection near the equator at the base of the convection zone. Whether such a latitudinal gap around the equator occurs on rapid rotators is unknown \citep[see][for a discussion]{Senavci21}, but one would expect that stochastic effects (e.g. convection) can induce scatter around the minimum latitude of emergence.  We reckon, however, that this additional scatter will not affect the photometric variability to a large degree.

When comparing different rows in Fig.~\ref{fig:map_spots}, one interesting feature becomes apparent: the spot area coverage is larger for the free nesting mode than for the non-nested case and active longitude nesting. 
This is because the proximity of neighbouring magnetic flux elements leads to a high possibility of same-polarity encounters. Magnetic flux, which accumulates this way, leads to the formation of spots.  We note that the possibility of spontaneous spot formation via flux superposition has been detected in numerical simulations \citep{Kitiashvili2010}, but so far such a formation has not been observed on the Sun, probably due to its relatively low activity level and a small degree of nesting of solar magnetic features.

Figures~\ref{fig:LCs_no_nest}--\ref{fig:LCs_FN_99} show that nesting affects not only the amplitude of the variability, but also the morphology of the LCs (see also discussion in Sect.~\ref{lightcurves}), in parallel with the computations by \citet{Isik2020}. For example, in the case of AL nesting, dips in the LCs mainly occur each half of the stellar rotation period. \cite{Basri2018} have introduced the "single/double ratio" (SDR) metric, which provides information about the ratio of time a star spends in single- or double-dip modes (i.e. its LC shows one or two peaks per period, respectively). The SDR was proposed to be an effective metric for characterizing stellar LCs. 
SDR as well as other metrics might be  useful for testing the models and constraining the distribution of stellar magnetic features. Indeed,
Fig.~\ref{fig:histograms} shows that both AL- and FN-type modes of nesting can result in the same amplitude of variability, albeit at different nesting degrees. Any metric, such as $R_{var}$, which is used in the present work, do not take into account LC morphology, as they only measure the peak-to-peak variability. Using metrics sensitive to the morphology of the LCs will help to distinguish between various modes of nesting. This will be addressed in a forthcoming publication.

Finally, our calculations are based on the assumption that the size distribution of emerging spots is the same on solar-like stars and the Sun, i.e. it does not depend on the rotation rate and stellar activity level. Specifically, the source term in \cite{Jiang2011_1} used in this study (see Step 1 in Sect.~\ref{FEAT}) is based on the sunspot size distribution during solar cycle 22. This neglects that the spot size distribution might depend on the activity level \citep[see, e.g.,][]{SolankiandUnruh2004, Krivovaetal2021}. We note that increase of spot sizes in the source term of SFTM model would simultaneously amplify the variability of a star and make its LC more regular. Thus,  the change of distribution of emerging spots might be another mechanism capable of explaining {\it Kepler} observations (together with nesting investigated in this study).  Since the lifetime of sunspots depends on their sizes, a change in the size distribution of spots would also lead to a change in their lifetimes, which would have a very direct effect on the variability amplitude and the LC statistics, with longer-living spots making the LC more regular. In that sense extending the lifetime of a spot should have a similar effect as stronger nesting. This is an important parameter whose influence will be studied in a future publication.

\section{Conclusions}\label{conclusions}

We coupled the model for the emergence and surface transport of magnetic flux in Sun-like stars \citep[][Paper I]{Isik2018} with a model for calculating stellar brightness variations \citep[partly based on the approach presented in][]{Nina1}. 
This allowed us to compute light curves of stars with rotation rates between 1 and 8\rotrate{} as they would be observed by \textit{Kepler} at different inclination angles. Following up on the findings of \cite{Isik2018} and \cite{Isik2020}, we investigated the impact of active-region nesting on the light curves and, in particular, the amplitude of the variability.%

We compared the output of our model to the observed variabilities of \textit{Kepler} stars in the temperature range 5500--6000K. In particular, we aimed at explaining the dependence of the amplitude of the variability on the rotation rate. Recently, \cite{Isik2020} showed that the model without nesting underestimates the variability of stars with known near-solar rotation periods \citep[see also][]{Timo2020}.  We found that the same is true for stars rotating faster than the Sun. Our runs without nesting dramatically underestimate the stellar variability at all rotation rates.

We showed that the observed dependence of {\it Kepler} variabilities on the rotation period, for stars with detected rotation periods, can be explained by an increase of nesting degree with the rotation rate, in parallel with increasing activity level. As both modes of nesting used in this work  led to similar levels of variability for stars with different rotation rates, we plan to further investigate the use of metrics that consider LC morphologies (instead of the peak-to-peak variability), to retrieve more information regarding the surface magnetic activity of stars. Additionally, we plan to include the effects of different rotation-activity relationships, cycle length and of stellar fundamental parameters (i.e. effective temperature and metallicities), on the variability. 
The applications of the FEAT model extend beyond stellar photometric variability, which is presented in this work. The model has been adapted to study the astrometric jitter introduced by stellar magnetic activity \citep{Sowmya2021_1, Sowmya2022} and Doppler imaging \citep{Senavci21}. It can also be used to study the magnetic contamination of high and low-resolution transmission spectra 
\citep[see, e.g.][]{Rackham1,Rackham2,Dravins2021, SAG21}.


\begin{acknowledgements}
The research leading to this paper has received funding from the European Research Council under the European Union’s Horizon 2020 research and innovation program (grant agreement No. 715947).

\end{acknowledgements}

\bibliographystyle{aa}
\bibliography{bib}



\end{document}